\documentclass[12pt,a4paper,english]{article}
\usepackage[T1]{fontenc}
\usepackage[latin1]{inputenc}
\usepackage{float}
\usepackage{amssymb}

\makeatletter

\providecommand{\tabularnewline}{\\}

\usepackage{float}
\usepackage{graphicx}
\usepackage[T1]{fontenc}

\newcommand{\beq}{\begin{equation}}
\newcommand{\eeq}{\end{equation}}

\newcommand{\ZZ}{\mathbb{Z}}

\newcommand{\bra}[1]{\langle #1 |}
\newcommand{\ket}[1]{|#1 \rangle}

\usepackage{amsfonts}
\usepackage{latexsym}

\usepackage{babel}
\makeatother
\begin{document}

\title{{\normalsize \begin{flushright} \vspace{-3cm} KCL-MTH-04-08\\
ITP--Budapest Report No. 611\\
hep-th/0406139
\end{flushright}\vspace{1cm}}
A nonperturbative study of phase transitions in the multi-frequency
sine-Gordon model}

\author{G\'{a}bor Zsolt T\'{o}th}

\maketitle
\begin{center}
\vspace{-0.6cm}
\textsl{Mathematics Department,}\\
 \textsl{King's College London, Strand, London WC2R 2LS, U.K.\footnote{until 
31th October 2004}}\\
\vspace{0.4cm}
\textsl{HAS Theoretical Physics Research Group,}\\
 \textsl{Institute for Theoretical Physics, E\"otv\"os University,}\\
 \textsl{H-1117 Budapest, P\'{a}zm\'{a}ny P\'{e}ter S\'{e}t\'{a}ny
1/A, Hungary}\\
\end{center}

\begin{center}\texttt{tgzs@ludens.elte.hu}\end{center}

\begin{abstract}
The phase spaces of the two- and three-frequency sine-Gordon models
are examined in the framework of truncated conformal space approach.
The focus is mainly on a tricritical point in the phase space of the
three-frequency model. We give substantial evidence that this point
exists. We also find the critical line in the phase space and present
TCSA data showing the change of the spectrum on the critical line
as the tricritical endpoint is approached. We find a few points of
the line of first order transition as well. 
\end{abstract}

\section{Introduction}

The sine-Gordon model in $(1+1)$ dimensions has attracted interest
long time ago for the reason that it appears in several areas of physics,
nevertheless it is an integrable field theory, that can be used to
study non-perturbative quantum field theory phenomena. The areas of
application include statistical mechanics of one-dimensional quantum
spin chains and nonlinear optics among many others --- see the introduction
of \cite{DM} for a representative list with references.

In the present paper we investigate a non-integrable extension of
the sine-Gordon model called multi-frequency sine-Gordon model, in
which the scalar potential consists of several cosine terms with different
frequencies. It is suggested in \cite{DM} that this model can be
used to give more refined approximation to some of the physical situations
where the ordinary sine-Gordon model can be used. A feature of the
multi-frequency model that is new compared to the usual sine-Gordon
model is --- apart form non-integrability --- that phase transition
can occur as the coupling constants are tuned. We concentrate our
attention to this property. It should be noted that such a phase transition
is related to the evolution of the particle spectrum of the theory
as the coupling constants vary, and we shall use the massgap and other
characteristics of the energy spectrum as order parameter.

Our investigation is a continuation of the work done on the double-frequency
case in \cite{BPTW}. We use the truncated conformal space approach
(TCSA), which is a non-perturbative numerical method to compute the
lowest levels of the spectrum of the Hamiltonian operator of perturbed
conformal field theories in finite volume. The applicability and reliability
of this method was thoroughly investigated in \cite{BPTW}, and it
was shown that the existence, nature and location of the phase transition
can be established by this method, although rather large number of
dimensions are needed for satisfactory precision. In particular, the
existence and location of an Ising type transition was established
in the double-frequency model (DSG) for the ratio $1/2$ of the frequencies,
verifying a prediction by \cite{DM} based on perturbation theory
and classical arguments. We extend these investigations to the ratio
$1/3$ and to the three-frequency model (at the ratio $1/2/3$ of
the frequencies), in which a tricritical point and first order transition
are expected to be found. The numerical nature of the TCSA makes it
necessary to choose specific values for the frequencies.

After introducing the multi-frequency sine-Gordon model and describing
basic properties of it in Section \ref{sec: ketto}, we briefly review
the TCSA framework for the model (which is the framework described
in \cite{BPTW} adapted to the multi-frequency model) in Section \ref{sec: harom}.
In Section \ref{sec: negy} we give a description of the phase structure
of the classical two- and three-frequency model, which serves as a
reference for the investigations in quantum theory. The $n$-frequency
case is also considered briefly. Exact and elementary analytic methods
can be applied to the classical case, and the results are more general
than in the quantum case. Section \ref{sec: ot} is devoted to theoretical
considerations on the signatures of 1st and 2nd order phase transitions
in the framework of perturbed conformal field theory in finite volume.
Most of these considerations, which are necessary for the evaluation
of the TCSA data, can also be found in \cite{BPTW}. Finally, in Section
\ref{sec: hat} and \ref{sec: het} we present the results we obtained
by TCSA on the phase structure of the two- and three-frequency model,
which are the main results in the paper. A summary of the results
and conclusions are given in Section \ref{sec:Conclusions}.

\section{\label{sec: ketto}The multi-frequency sine-Gordon model}

\subsection{The definition and basic properties of the model}

The action of the multi-frequency sine-Gordon model (MSG) is

\[
\mathcal{A}_{MSG}=\int\mathrm{d}t\int\mathrm{d}x\,\left(\frac{1}{2}\partial_{\mu}\Phi\partial^{\mu}\Phi-V(\Phi)\right),\]
 where \[
V(\Phi)=\sum_{i}^{n}\mu_{i}\cos(\beta_{i}\Phi+\delta_{i})\]
 is the potential, which contains $n$ cosine terms. $\Phi$ is a
real scalar field defined on the two-dimensional Minkowski space $\mathbb{R}^{1+1}$,
$\beta_{i}\in\mathbb{R}$ are the frequencies, $\beta_{i}\ne\beta_{j}$
if $i\ne j$, $\mu_{i}$ are the coupling constants (of dimension
mass$^{2}$ at the classical level) and $\delta_{i}\in\mathbb{R}$
are the phases in the terms of the potential.

Two cases can be distinguished according to the periodicity properties
of the potential. The first one is the rational case, when $V(\Phi)$
is a trigonometric polynomial: the ratios of the frequencies $\beta_{i}$
are rational and the potential is periodic. Let the period of the
potential be $2\pi r$ in this case. The target space of the field
$\Phi$ can be compactified: \[
\Phi\equiv\Phi+2kr\pi,\]
 where $k\in\mathbb{N}$ can be chosen arbitrarily. The model obtained
in this way is called the $k$-folded MSG. The well-known classical
sine-Gordon model corresponds to $n=1$, $k=1$.

The other case is the irrational one, when the potential is not periodic
and no such folding can be made. The irrational case is much more
complicated than the rational one, so we restrict our attention to
the rational case in the present paper. We remark here only that although
$V(\Phi)$ always has a finite infimum, it does not necessarily admit
an absolute minimum. The potential $V(\Phi)$ can always be written
uniquely as a sum $V(\Phi)=V_{1}(\Phi)+V_{2}(\Phi)+...+V_{k}(\Phi)$,
where the terms $V_{1},...,V_{k}$ are periodic but any sum of any
of these terms is not periodic. $V(\Phi)$ has an absolute minimum
if and only if $V_{1}(\Phi),...,V_{k}(\Phi)$ have a common absolute
minimum. This occurs for special choice of the $\delta_{i}$, if the
values of $\beta_{i}$ are given. In particular, if $\beta_{i}/\beta_{j}$
are irrational for all $i\ne j$ and $\mu_{i}<0$ for all $i$, then
$V(\Phi)$ has a absolute minimum if and only if $\frac{\delta_{i}}{\beta_{i}}-\frac{\delta_{j}}{\beta_{j}}=\frac{2\pi b_{i}}{\beta_{i}}-\frac{2\pi b_{j}}{\beta_{j}}$
is satisfied with some numbers $b_{i}\in\ZZ$, which is equivalent
to the case $\delta_{i}=0$ for all $i$. See and \cite{DM} and \cite{BPTW}
for further remarks on the irrational case.

At the quantum level the theory can be regarded as a perturbed conformal
field theory: \[
\mathcal{A}_{MSG}=\mathcal{A}_{CFT}+\mathcal{A}_{pert},\]
 where\[
\mathcal{A}_{CFT}=\int\mathrm{d}t\int\mathrm{d}x\,\frac{1}{2}\partial_{\mu}\Phi\partial^{\mu}\Phi,\]
 which is the action of the free scalar particle of zero mass, and\[
\mathcal{A}_{pert}=\int\mathrm{d}t\int\mathrm{d}x\,(-V(\Phi))=-\frac{1}{2}\int\mathrm{d}t\int\mathrm{d}x\,\sum_{i=1}^{n}(\mu_{i}e^{i\delta_{i}}V_{\beta_{i}}+\mu_{i}e^{-i\delta_{i}}V_{-\beta_{i}}),\]
 where $V_{\omega}$ denotes the vertex operator \[
V_{\omega}=:e^{i\omega\Phi}:,\]
 which is a primary field with conformal dimensions \[
\Delta_{\omega}^{\pm}=\Delta_{\omega}=\frac{\omega^{2}}{8\pi}\]
 in the unperturbed (conformal) field theory. The upper index $\pm$
corresponds to the left/right conformal algebra and : : denotes the
conformal normal ordering. The dimensions of the couplings at the
quantum level are \[
[\mu_{i}]=(mass)^{2-2\Delta_{i}},\qquad\Delta_{i}\equiv\Delta_{\beta_{i}}.\]
 The perturbing operators are relevant only if \begin{equation}
\beta_{i}^{2}<8\pi,\label{relevance}\end{equation}
 we restrict ourselves to this case. We also assume that \[
\beta_{i}^{2}<4\pi,\]
 which is a necessary and sufficient condition for the model to be
free from ultraviolet divergencies in the perturbed conformal field
theory framework \cite{Kl-M,DM,BPTW}.

The model has a massgap in general, and it is clear that phase transitions
occur in the classical version of the model as the coupling constants
are tuned (assuming that $n>1$). It is also expected that there are
topologically charged solutions/states in the model \cite{DM}. In
the present paper we shall investigate the sector with zero topological
charge, which is sufficient for our purposes. We also restrict ourselves
to 1-folded models ($k=1$), as it is natural to expect that in infinite
volume a folding number $k\ne1$ results simply in a $k$-fold multiplication
of the spectrum corresponding to $k=1$.

\section{\label{sec: harom}The Truncated Conformal Space Approach}

The matrix elements of the total Hamiltonian operator $H=H_{CFT}+H_{pert}$
can be calculated explicitly (and exactly) between any two elements
of the Hilbert space $\mathcal{H}_{CFT}$ of the conformal field theory.
Restricting to a finite dimensional subspace $\mathcal{H}_{0}$ of
$\mathcal{H}_{CFT}$ by introducing an upper conformal energy cutoff
$H$ becomes a finite matrix, which can be diagonalized numerically
to get an approximation of the spectrum. As this description shows,
TCSA is a variational method.

The $k$-folded free boson as a conformal field theory contains the
following primary fields:

\[
V_{p,\bar{p}}(z,\bar{z})=:\exp[ip\phi_{CFT}(z)+i\bar{p}\bar{\phi}_{CFT}(\bar{z})]:\]
 with conformal dimensions $\Delta^{+}=\frac{p^{2}}{8\pi}$, $\Delta^{-}=\frac{\bar{p}^{2}}{8\pi}$,
where $p=\frac{n}{r}+2\pi rm$, $\bar{p}=\frac{n}{r}-2\pi rm$, $n,m\in\ZZ$,\[
\Phi_{CFT}(x,t)=\phi_{CFT}(x-t)+\bar{\phi}_{CFT}(x+t).\]
 $\mathcal{H}_{CFT}$ is spanned by the states $\ket{p,\bar{p}}=\lim_{z,\bar{z}\rightarrow0}V_{p,\bar{p}}(z,\bar{z})\ket{0}$
($\ket{0,0}\equiv\ket{0}$) and $a_{n_{1}}...\bar{a}_{m_{1}...}\ket{p,\bar{p}}$,
where $a_{n_{i}}$ and $\bar{a}_{m_{i}}$ are creating operators of
Fourier modes on the {}``conformal plane''. The conformal generators
$L_{0}$ and $\bar{L}_{0}$ are diagonal in this basis. The basis of $\mathcal{H}_{0}$
is obtained by taking those elements $\ket{v}$ of the basis above
which satisfy the truncation condition

\[
\frac{\bra{v}\frac{L}{2\pi}H_{CFT}\ket{v}}{\bra{v}\ket{v}}<e_{cut}.\]
 $e_{cut}$ is the dimensionless upper conformal energy cutoff and
$L$ is the volume of space. We restrict ourselves to the sector with
zero topological charge ($p=\bar{p}$), this being the sector containing
the ground state(s) and the relevant information for the problem treated
in this work. In the sector containing the ground state(s) the Lorentz
spin of all the states is also zero, so we also impose the condition
$(L_{0}-\bar{L}_{0})\ket{v}=0$. (The operator $L_{0}-\bar{L}_{0}$
commutes with $H$ and $H_{CFT}$ as well). We remark that $e_{cut}$
serves as an ultraviolet cutoff.

The matrix elements of $H$ between two elements $\ket{a}$ and $\ket{b}$
of the basis of $\mathcal{H}_{CFT}$ above are given by \begin{equation}
\left(\frac{H}{M}\right)_{ab}=\frac{2\pi}{l}\left(L_{0}+\bar{L}_{0}-\frac{c}{12}\right)_{ab}\label{TCSA Hamilton}\end{equation}
 \[
+\frac{2\pi}{l}\sum_{j=1}^{n}sgn(\mu_{j})\kappa_{j}\left(\frac{M_{j}}{M}\right)^{x_{j}}\frac{l^{x_{j}}}{2(2\pi)^{x_{j}-1}}e^{i\delta_{j}}(V_{\beta_{j},\beta_{j}}(1,1))_{ab}\delta_{\Delta_{a}-\bar{\Delta}_{a},\Delta_{b}-\bar{\Delta}_{b}}\]
 \[
+\frac{2\pi}{l}\sum_{j=1}^{n}sgn(\mu_{j})\kappa_{j}\left(\frac{M_{j}}{M}\right)^{x_{j}}\frac{l^{x_{j}}}{2(2\pi)^{x_{j}-1}}e^{-i\delta_{j}}(V_{-\beta_{j},-\beta_{j}}(1,1))_{ab}\delta_{\Delta_{a}-\bar{\Delta}_{a},\Delta_{b}-\bar{\Delta}_{b}},\]
 where $M$ is a mass scale of the theory given below, $l=LM$ is
the dimensionless volume, $x_{j}=2-2\Delta_{j}$, $\Delta_{j}\equiv\Delta_{\beta_{j}}$;
$\Delta_{a}$, $\bar{\Delta}_{a}$, $\Delta_{b}$, $\bar{\Delta}_{b}$
are the conformal weights of the states $\ket{a}$ and $\ket{b}$,
$c$ is the central charge of the conformal theory ($c=1$ in the
present case), and we have made a replacement corresponding to \[
|\mu_{j}|=\kappa_{j}M_{j}^{x_{j}}.\]
 The {}``interpolating'' mass scale $M$ is \[
M=\sum_{j}\eta_{j}M_{j},\]
 where \begin{equation}
\eta_{j}=\frac{|\mu_{j}|^{1/x_{j}}}{\sum_{i}|\mu_{i}|^{1/x_{i}}}\label{dless coupling}\end{equation}
 are the dimensionless coupling constants (of which only $n-1$ are
independent). (\ref{dless coupling}) implies that $\eta_{j}\in[0,1]$,
$\sum_{j}\eta_{j}=1$. $M$ depends smoothly on the $\eta_{j}$-s.
The precise expression for $\kappa$ is not essential for our problem,
we need only that $\kappa$ depends on $\Delta$ only and that it
is dimensionless. Following \cite{BPTW} we used the formula of \cite{Zam}:\[
\kappa_{j}=\frac{2\Gamma(\Delta_{j})}{\pi\Gamma(1-\Delta_{j})}\left(\frac{\sqrt{\pi}\Gamma(\frac{1}{x_{j}})}{2\Gamma(\frac{\Delta_{j}}{x_{j}})}\right)^{x_{j}}.\]

The formula (\ref{TCSA Hamilton}) is written in terms of dimensionless
quantities, and the volume $(l)$ dependence of $H/M$ is also explicit.
It is clear from (\ref{TCSA Hamilton}) that TCSA gives an exact result
if $l\rightarrow0$ (assuming (\ref{relevance})) and this limit of
the theory is the conformal theory (the massless free boson), and
the accuracy of the TCSA spectrum decreases at fixed $e_{cut}$ as
$l\rightarrow\infty$. For very large values of $l$ the $l$-dependence
of the spectrum of the TCSA Hamiltonian is power-like, as it is determined
directly by the $l^{x_{j}-1}$ coefficients in (\ref{TCSA Hamilton}).
The TCSA Hamiltonian cannot be considered as good approximation for
these values of $l$. Increasing the value of $e_{cut}$ generally
increases accuracy, and the TCSA is expected to give better approximation
for the lower lying energy levels than for the higher ones.

We denote the (dimensionless) energy levels of $H/M$ in volume $l$
by $e_{i}(l)$, $i=0,1,2,...$, and $e_{0}\leq e_{1}\leq e_{2}\leq...$
if not stated otherwise. We shall draw conclusions about the spectrum
at $l=\infty$ from the behaviour of the functions $e_{i}(l)$ for
low values of $i$ and moderately large values of $l$.

We refer to \cite{BPTW,key-21,Yurov-Zam} for more details on the
TCSA framework.

\section{\label{sec: negy}Phase structure in the classical limit }

\subsection{Phase structure of the two-frequency model in the classical limit}

The Lagrangian density takes the following form in the two-frequency
case: \[
\mathcal{L}=\frac{1}{2}\partial_{\mu}\Phi\partial^{\mu}\Phi-\mu\cos(\beta\Phi)-\lambda\cos(\alpha\Phi+\delta),\]
 where \[
\frac{\beta}{\alpha}=\frac{n}{m}\ne1,\]
 $n$ and $m$ are coprimes (and the folding number equals to one).

Proposition 1: Assume that $\mu,\lambda\ne0$. Then the following
three cases can be distinguished:

a.) If the function $V(\Phi)$ is symmetric with respect to the reflection
$\Phi\mapsto2\Phi_{0}-\Phi$, where $\Phi_{0}$ is a suitable constant, and
$n,m>1$, then $V$ has two absolute minima, which are mapped into
each other by the reflection. We remark that it depends on the value
of $\delta$ whether $V$ has this symmetry or not, and it is easy
to give a criterion for the existence of this symmetry in terms of
$n,m$ and $\delta$.

b.) If $V$ is symmetric with respect to a reflection as in case a.,
but $n=1$ or $m=1$, then $V$ has one or two absolute minima depending
on the values of $\mu$ and $\lambda$. In this case, assuming that
$n=1$, $V$ can be brought to the form \[
V(\Phi)=-|\mu|\cos(\beta\Phi)+|\lambda|\cos(m\beta\Phi)\]
 by an appropriate shift of $\Phi$. $V$ has two absolute minima
if $|\lambda/\mu|>1/m^{2}$, and one absolute minimum if $|\lambda/\mu|\leq1/m^{2}$.
The two absolute minima are mapped into each other by the reflection.
The second derivative of $V$ is nonzero at the minima if $|\lambda/\mu|\ne1/m^{2}$,
but it is zero if $|\lambda/\mu|=1/m^{2}$. In the latter case, the
fourth derivative of $V$ at the minimum is nonzero. The two minima
of $V$ merge and the value of the second derivatives of $V$ at the
two minima tends to zero as $|\lambda/\mu|$ approaches $1/m^{2}$
from above.

c.) If $V$ does not satisfy the requirements of a) and b), then $V$
has a single absolute minimum.

We omit the proof of this proposition, which is elementary, although
long and not completely trivial because of the arbitrariness of $n$
and $m$.

If $\mu$ or $\lambda$ equals to zero, then $V$ is periodic and
has $m$ or $n$ absolute minima, respectively. See \cite{key-21}
for a detailed investigation of these (integrable) limiting cases.

The phases of the classical model are determined by the behaviour
of absolute minima of $V(\Phi)$ as the value of the coupling constants
vary. In particular, Proposition 1 implies that the phase structure
of the two-frequency model is the following:

The model exhibits an Ising-type second order phase transition at
the critical value \[
\eta_{c}=\frac{m}{1+m}\]
 of the dimensionless coupling constant $\eta=\sqrt{|\mu|}/(\sqrt{|\mu|}+\sqrt{|\lambda|})$
if $n=1$ and $V$ has the $\ZZ_{2}$-symmetry introduced above. This
critical point separates two massive phases with unbroken and spontaneously
broken $\ZZ_{2}$-symmetry. Equivalent statement can be made if $m=1$.
If $V$ is not symmetric, then there is only one massive phase with
nondegenerate ground state. If $m,n\ne1$ and $V$ is symmetric, then
there is one massive phase with doubly degenerate ground state (i.e.
the reflection symmetry is spontaneously broken). In the limiting
cases $\eta=0$ and $\eta=1$ the model is massive and has spontaneously
(and completely) broken $\ZZ_{n}$ or $\ZZ_{m}$ symmetry.

\subsection{\label{sec 4.2}Phase structure of the three-frequency model in the
classical limit}

A complete description of the behaviour of the absolute minima of
$V$ for all values of the parameters becomes excessively difficult
in the three- and higher-frequency cases, so we restrict our attention
to particular values. The potential in the three-frequency case is
\begin{equation}
V(\Phi)=\mu_{1}\cos(\beta_{1}\Phi)+\mu_{2}\cos(\beta_{2}\Phi+\delta_{2})+\mu_{3}\cos(\beta_{3}\Phi+\delta_{3}).\label{eq: 3freq pot}\end{equation}
 We choose the frequency ratios $3:2:1$, i.e. \[
\beta_{1}=\beta,\qquad\beta_{2}=\frac{2}{3}\beta,\qquad\beta_{3}=\frac{1}{3}\beta.\]
 This three-frequency model has a tricritical point if and only if
$\delta_{2}=\delta_{3}=0$ (and also in a few equivalent cases), in
this case $V$ is symmetric with respect to the reflection $\Phi\mapsto-\Phi$.
In the tricritical point the absolute minimum of $V$ can be located
only at $0$ or $\pi$. The two cases are equivalent, we consider
the case when the location of the absolute minimum is $0$. The tricritical
point in this case is located at\[
\frac{\mu_{1}}{\mu_{2}}=-\frac{1}{6},\qquad\frac{\mu_{1}}{\mu_{3}}=\frac{1}{15}.\]
 In this point $V^{(6)}(0)\ne0$. (The upper index $^{(6)}$ denotes
the sixth derivative with respect to $\Phi$.) $V^{(6)}(0)>0$ requires
$\mu_{1},\mu_{3}<0$ and $\mu_{2}>0$. We restrict ourselves to this
domain and to the values $\delta_{2}=\delta_{3}=0$.

\begin{figure}
\begin{center}\includegraphics[%
  scale=0.7]{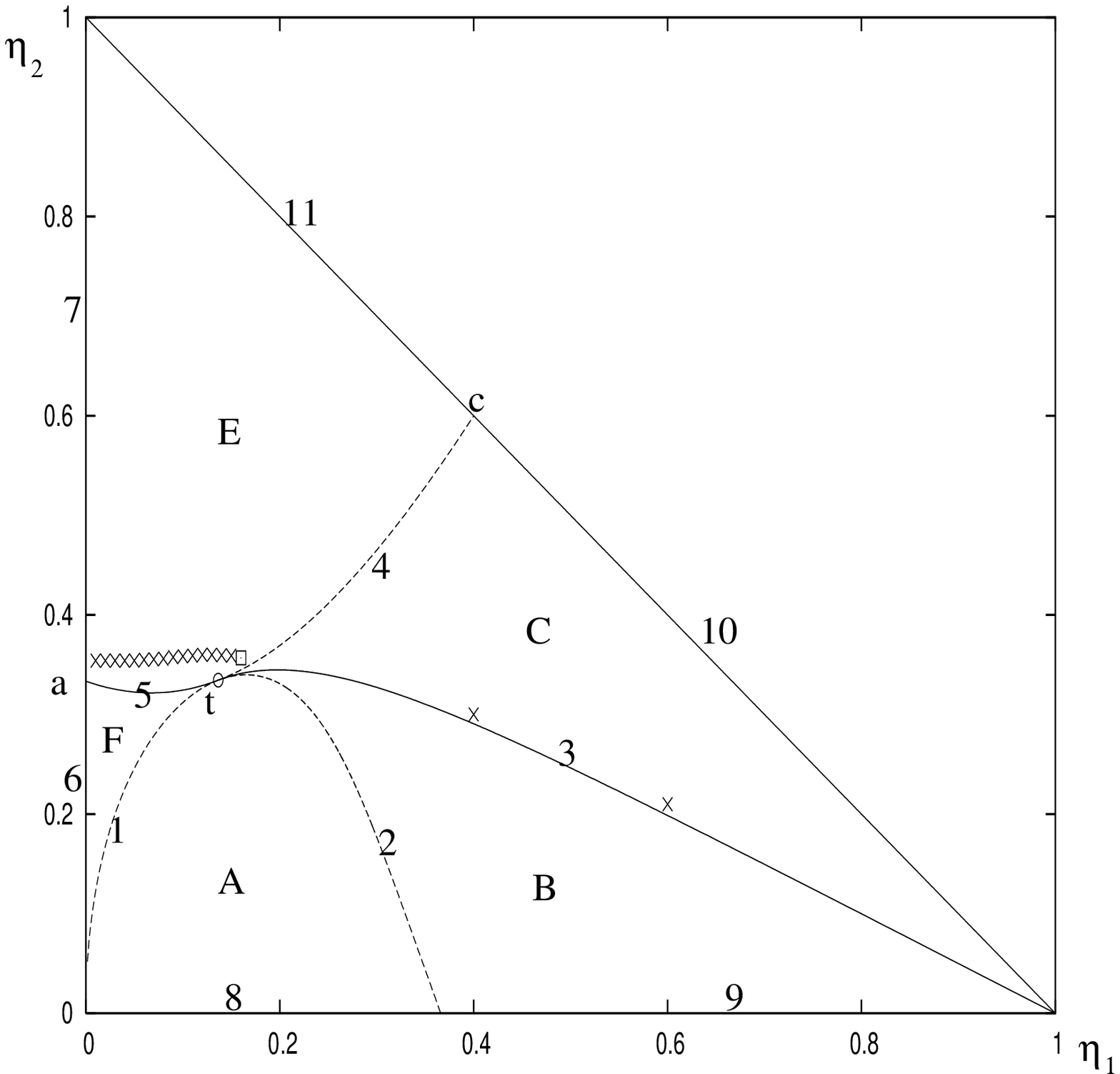}\end{center}

\caption{\label{threefreq phasediag}Phase diagram of the classical three-frequency
sine-Gordon model at $\beta_{1}/\beta_{2}/\beta_{3}=3/2/1$, $\delta_{1}=\delta_{2}=\delta_{3}=0$,
$\mu_{1},\mu_{3}<0$ and $\mu_{2}>0$. The crosses and the square correspond to certain
quantum theory values described in Section \ref{sec: het}.}
\end{figure}

\begin{figure}
\begin{center}\includegraphics[%
  height=0.17\textwidth]{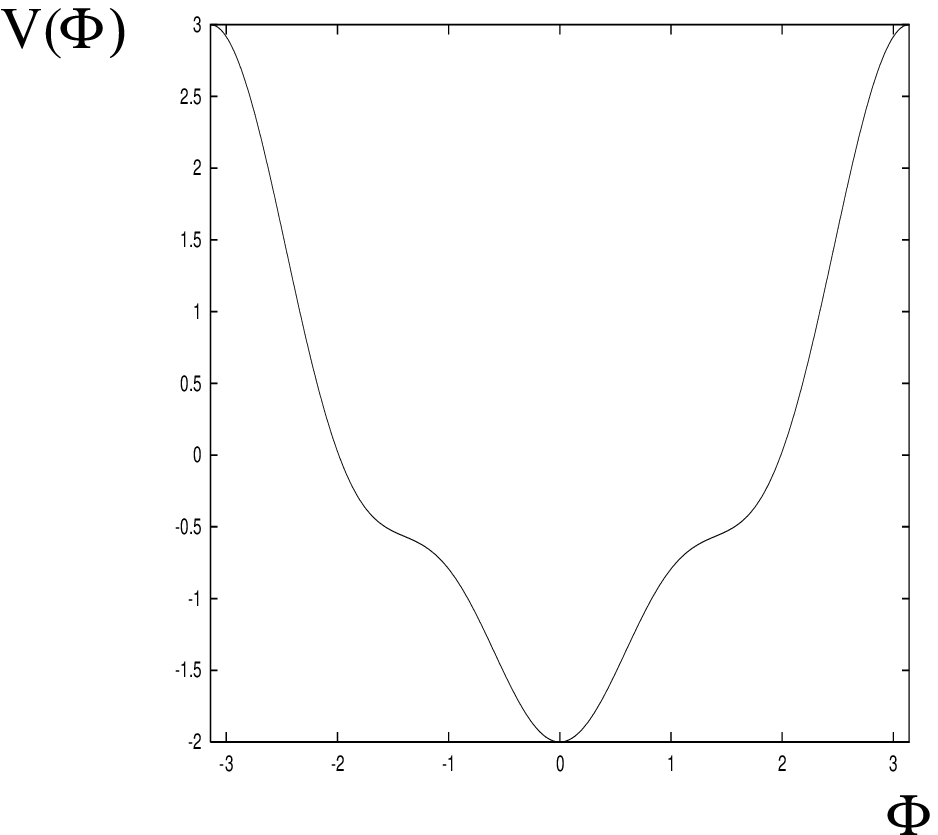}\includegraphics[%
  height=0.17\textwidth]{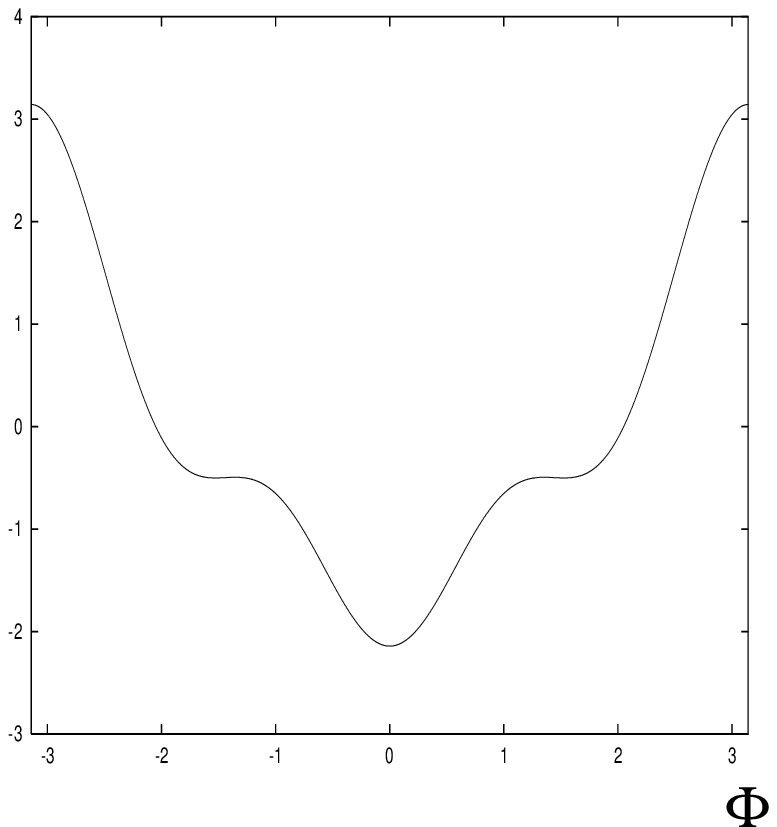}\includegraphics[%
  height=0.17\textwidth]{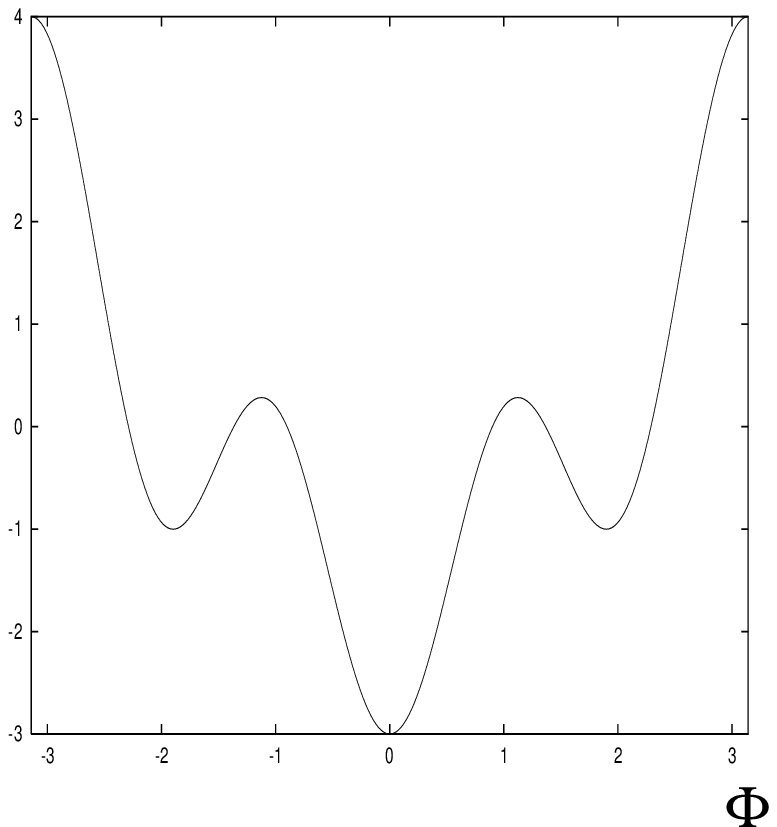}\includegraphics[%
  height=0.17\textwidth]{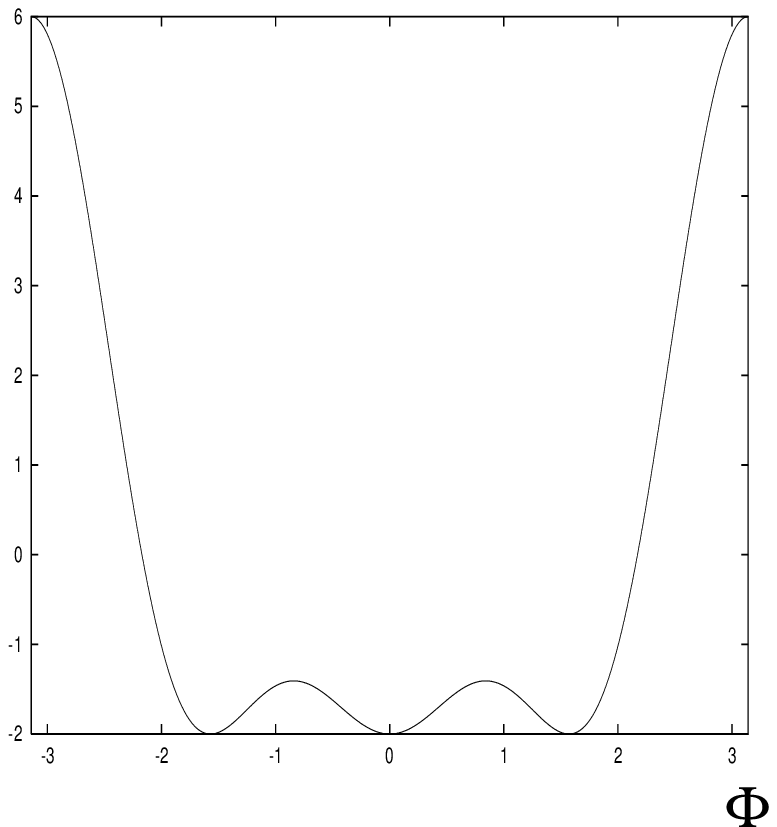}\end{center}

\vspace{-0.5cm} \hspace{3.7cm}1,6,8,A,F\hspace{1.5cm}2\hspace{2.2cm}B\hspace{2.3cm}3

\begin{center}\includegraphics[%
  height=0.17\textwidth]{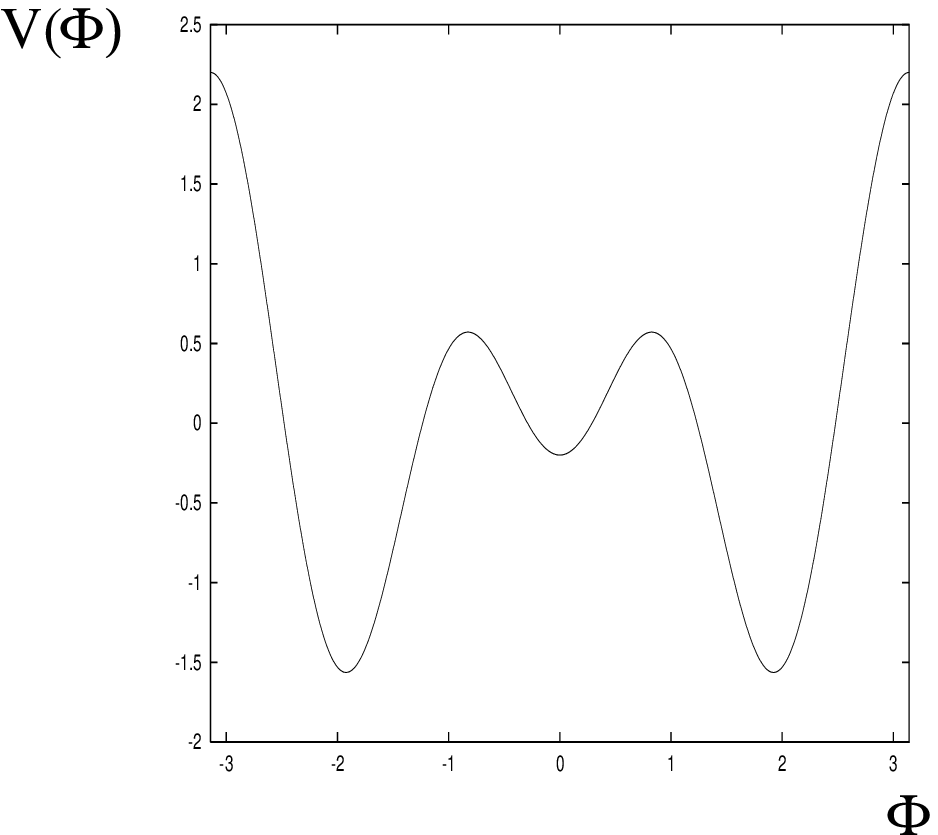}\includegraphics[%
  height=0.17\textwidth]{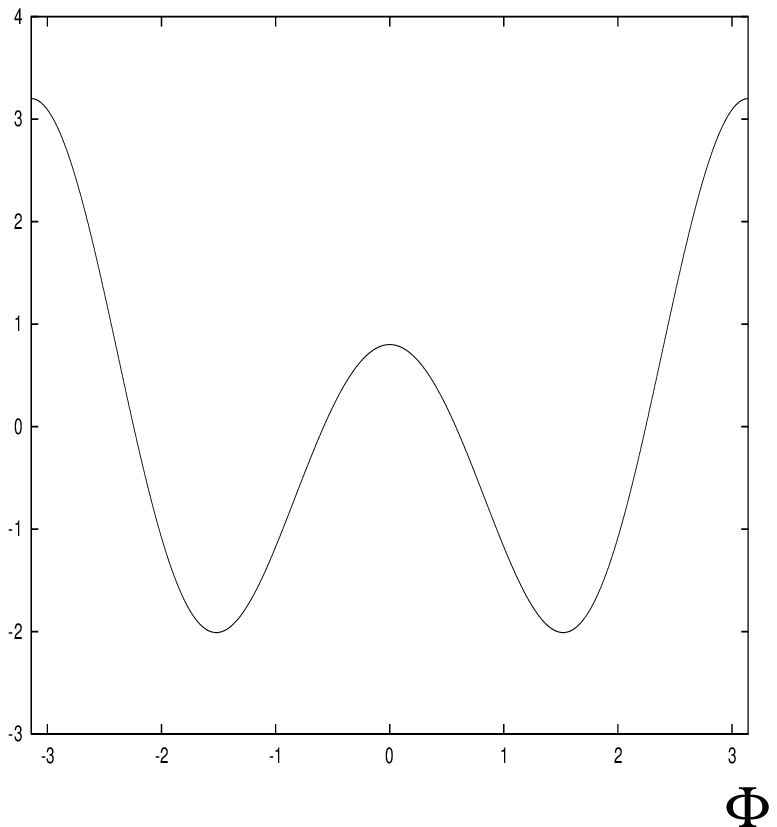}\includegraphics[%
  height=0.17\textwidth]{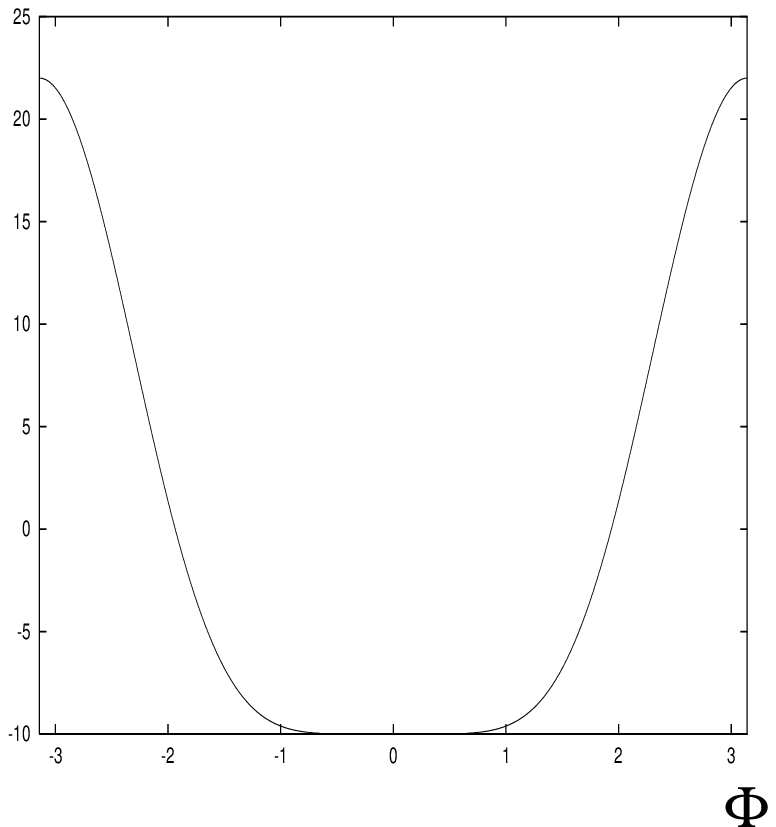}\end{center}

\vspace{-0.5cm} \hspace{5.7cm}C\hspace{1.8cm}4,E,7\hspace{1.6cm}a,5,t

\caption{\label{tpotfigs}Characteristic shapes of the potential }
\end{figure}

The phase diagram is shown in Figure \ref{threefreq phasediag}. The
points of the diagram correspond to the values of the pair $(\eta_{1},\eta_{2})$
of dimensionless parameters. The allowed values constitute the left
lower triangle, the straight line joining $(0,1)$ and $(1,0)$ corresponds
to $\eta_{3}=0$. The tricritical point is denoted by $t$, it is
located at \[
\left(\frac{1}{1+\sqrt{6}+\sqrt{15}},\frac{\sqrt{6}}{1+\sqrt{6}+\sqrt{15}}\right)\approx(0.1365,0.3345).\]
 At $t$ $V$ has one single and absolute minimum (at $\Phi=0$).
Phase transition occurs when the lines $5$ and $3$ shown in the
phase diagram are crossed. Second order Ising-type phase transition
occurs on $5$ and first order phase transition occurs on $3$. The
domain $A\cup B\cup F$ corresponds to a massive $\ZZ_{2}$-symmetric
phase (with unique ground state). The domain $E\cup C$ corresponds
to a massive phase with spontaneously broken $\ZZ_{2}$-symmetry.
Characteristic shapes of the potential in the various domains and
on the various lines of the phase diagram can be seen in Figure \ref{tpotfigs}.
Data applying to the quantum case are also shown in Figure \ref{threefreq phasediag},
they will be explained in subsequent sections.

\subsection{$n$-frequency model in the classical limit }

Let us take the $n$-frequency model with $\beta_{i}=i\beta$, $i=1..n$,
and $\delta_{i}=0$. In this case there exist unique values of $\mu_{i}/\mu_{1}$,
$i=2..n$ so that $V(x)$ has a single global minimum at $x=0$ and
has no other local minima, and $V''(0)=0$, $V''''(0)=0$, ... ,$V^{(2n)}(0)=0$
also hold. The values of $\mu_{i}/\mu_{1}$ are determined by the
latter equations. The point corresponding to these values of $\mu_{i}/\mu_{1}$
is an $n$-fold multicritical point in the phase space. The neigbourhood
of this multicritical point contains $m$-fold multicritical points
for any integer $0<m<n$.

These statements can be proved using well known properties of analytic
functions and the fact that $V$ is a trigonometric polynomial. We
omit the details of the proof.

\section{\label{sec: ot}Signatures of 1st and 2nd order phase transitions
in finite volume}

The considerations in this section apply to quantum field theory.

The behaviour of the spectrum is governed by the $l\rightarrow0$
limiting conformal field theory for small values of $l$, so $e_{n}(l)-e_{0}(l)\sim1/l$.
Massive phases in infinite volume are characterized by the existence
of a massgap and the behaviour $\lim_{l\rightarrow\infty}(e_{n}(l)-e_{0}(l))=C_{n}$,
where $C_{n}\geq0$ are constants. $C_{n}=0$ if $0\leq n\leq d$
and $C_{n}>0$ if $n>d$, if the ground state has $d$-fold degeneracy
in infinite volume. In a phase with spontaneously broken symmetry
the spectrum is degenerate in the $l\rightarrow\infty$ limit, in
finite volume the degeneracy is lifted (at least partially) due to
tunneling effects. The resulting energy split between the degenerate
vacua vanishes exponentially as $l\rightarrow\infty$.

In the critical points (in infinite volume) the massgap vanishes and
the Hilbert space contains a sector that corresponds to the conformal
field theory specifying the universality class of the critical point.
We consider this sector in the following discussion. In finite but
large volume near the critical point this sector of the theory can
be regarded as the $l\rightarrow\infty$ limiting conformal theory
perturbed by some irrelevant and relevant operators. The corresponding
TCSA Hamiltonian takes the (generic) form \begin{equation}
H=\frac{2\pi}{L}\left((L_{0})_{IR}+(\bar{L}_{0})_{IR}-\frac{c_{IR}}{12}+\sum_{\psi}\frac{g_{\psi}L^{2-2\Delta_{\psi}}}{(2\pi)^{1-2\Delta_{\psi}}}\psi(1,1)\right),\label{IR Hamilton}\end{equation}
 where the $\psi$ are the perturbing fields. This picture gives the
following volume dependence of energy levels (in the first order of
conformal perturbation theory):\begin{equation}
e_{\Psi}(l)-e_{0}(l)=\frac{2\pi}{l}(\Delta_{IR,\Psi}^{+}+\Delta_{IR,\Psi}^{-})+\sum_{\psi}A_{\Psi}^{\psi}l^{1-2\Delta_{\psi}},\label{l fugges}\end{equation}
 where $\Delta_{IR,\Psi}^{+}$ and $\Delta_{IR,\Psi}^{-}$ are the
conformal weights of the state $\Psi$ in the $l\rightarrow\infty$
limiting CFT, $A_{\Psi}^{\psi}$ are constants that also depend on
the particular energy eigenstate $\Psi$. The presence of irrelevant
perturbations ($1-2\Delta_{\psi}<-1$) is due to the finiteness of
the volume, whereas the presence of the relevant perturbations ($1-2\Delta_{\psi}>-1$)
is caused by the deviation of the control parameters from the critical
value and by the UV cutoff.

The location of the critical points can be determined using the criterion
of vanishing massgap. A more precise method that also allows the determination
of the universality class of the critical point (i.e. the $l\rightarrow\infty$
limiting CFT) is the following: We make an assumption that the critical
point is in a certain universality class. This assumption predicts
the set of $\psi$-s, the values of $\Delta_{IR,\Psi}^{+}$ and $\Delta_{IR,\Psi}^{-}$,
and the values of the $\Delta_{\psi}$-s in (\ref{l fugges}). We
take leading terms of the series on the r.h.s. of (\ref{l fugges})
and determine the value of the $(\Delta_{IR,\Psi}^{+}+\Delta_{IR,\Psi}^{-})$-s
and of the $A_{\Psi}^{\psi}$-s by fitting to the TCSA energy data.
The magnitude of the $A_{\Psi}^{\psi}$-s corresponding to the relevant
perturbations measures the deviation from the critical point, so if
the assumption on the universality class is right, then by tuning
the coupling constants one should be able to find a (critical) value
of the control parameter at which these $A_{\Psi}^{\psi}$-s are small,
the TCSA data are described well by (\ref{l fugges}) (terms from
higher orders of perturbation theory can be included if necessary)
in a reasonably large interval of the values of $l$, and the values
of the $(\Delta_{IR,\Psi}^{+}+\Delta_{IR,\Psi})$-s obtained from
the TCSA data agree with the assumption with good precision. The interval
where (\ref{l fugges}) describes the TCSA data well is called the
scaling region. This region may be (and in fact is) different for
different energy levels. We remark that it is also possible to make
a theoretical prediction for $e_{0}(l)$, which allows to extract
$c_{IR}$ from the TCSA data for $e_{0}(l)$ in principle \cite{BPTW}. However,
experience (\cite{BPTW}) shows that the accuracy of the TCSA data is not sufficient to determine 
$c_{IR}$ precisely in
this way, so we do not attempt to extract $c_{IR}$ directly from the TCSA data in this paper.

In the classical case a first order phase transition occurs when the
absolute minimum of the potential becomes a relative minimum and a
previously relative minimum becomes absolute. In the quantum case
this phase transition is characterized by the presence of {}``runaway''
energy levels with asymptotic behaviour $e(l)\sim cl$ for large $l$
in the neighborhood of the transition point, where $c$ is a constant
that tends to zero as the transition point is approached. The multiplicity
of the ground state also changes as the transition point is passed
if the two phases have different symmetry properties. We remark that
{}``runaway'' energy levels are present in general whenever a model
has unstable vacua.

\section{\label{sec: hat}The phase diagram of the two-frequency model in
the case $\frac{\alpha}{\beta}=\frac{1}{3}$, $\delta=\frac{\pi}{3}$}

We assume that $\lambda,\mu>0$, and use the parameter \[
\tilde{\eta}=\frac{\lambda^{x_{\beta}}}{\mu^{x_{\alpha}}+\lambda^{x_{\beta}}}\]
 instead of $\eta$ in this case to conform with \cite{BPTW}.

The classical model with $\alpha/\beta=\frac{1}{3}$, $\delta=\frac{\pi}{3}$
exhibits an Ising type phase transition at $\tilde{\eta}=3^{4}/(1+3^{4})$.
\begin{figure}
\begin{center}\includegraphics[%
  width=0.30\linewidth,
  angle=270]{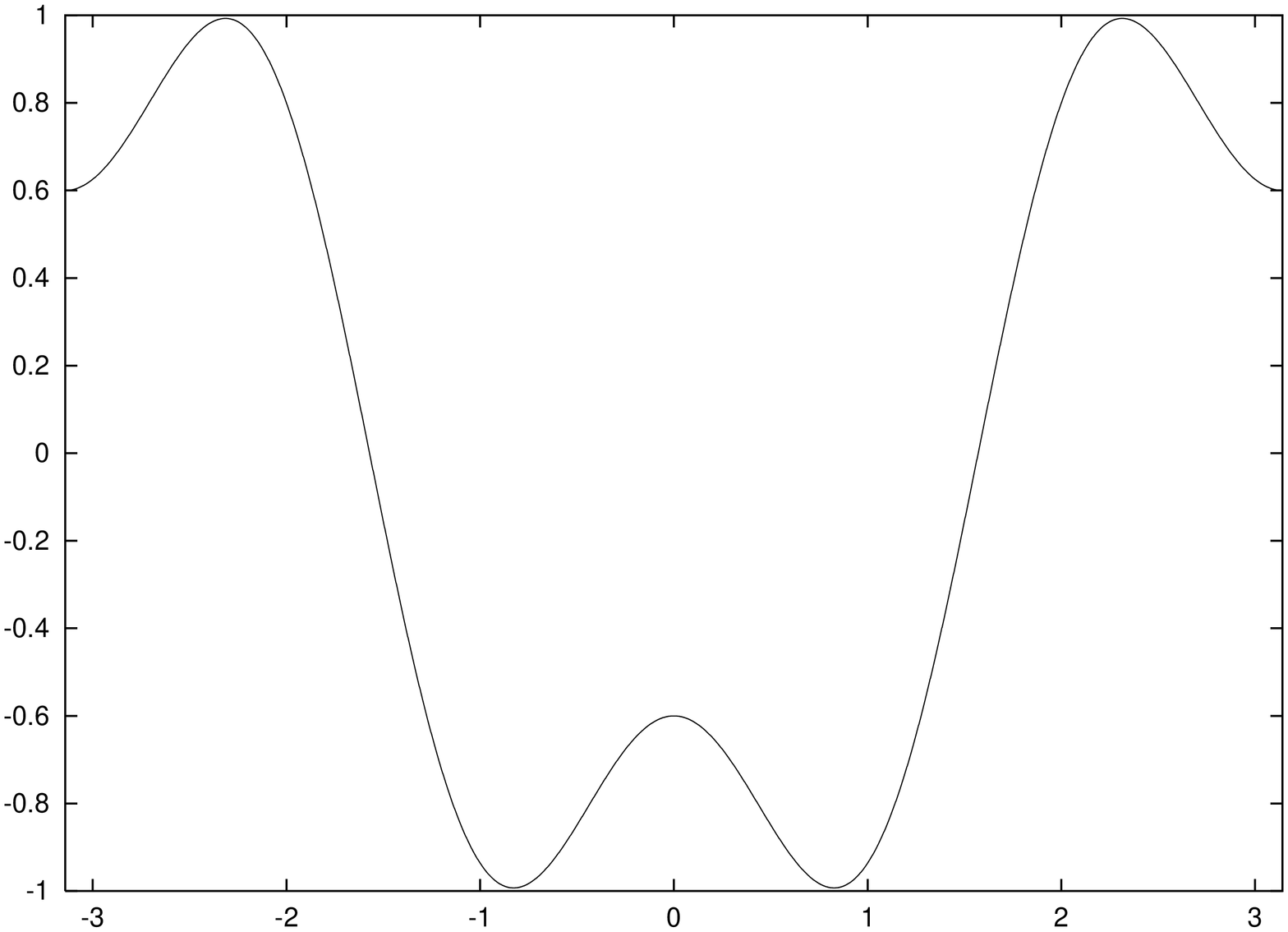}\includegraphics[%
  width=0.30\linewidth,
  angle=270]{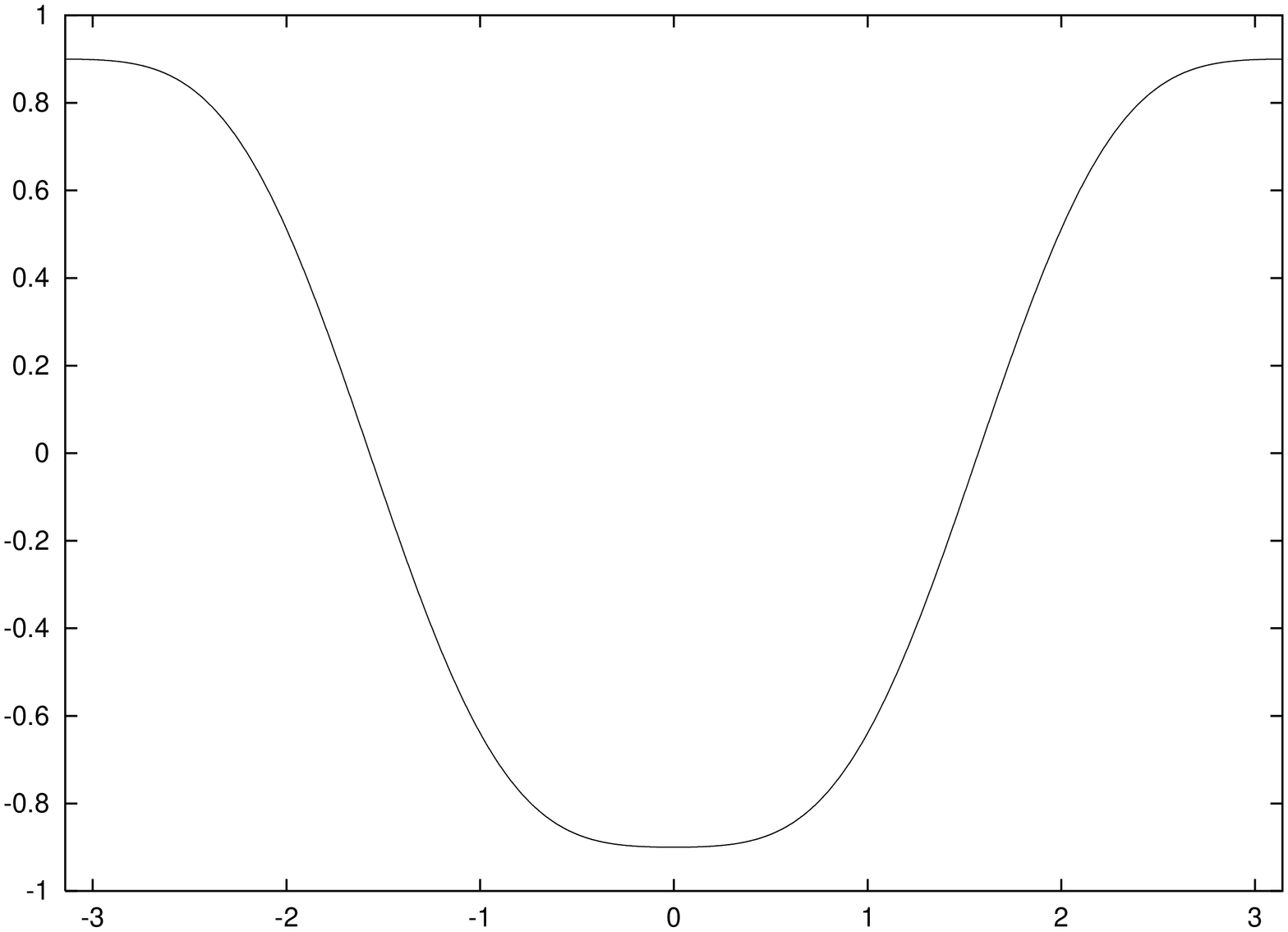}\end{center}

\hspace{4.5cm}a\hspace{7cm}b

\caption{\label{pot12}Typical shape of $V(\Phi)$ in the broken and in the
unbroken symmetry phase}
\end{figure}

Considering the quantum case we proceed along the lines of \cite{BPTW}
in this section. The numerical nature of the TCSA makes it necessary
to choose a finite number of values for $\beta$ and $l$ at which
calculations are done. One should choose as large values for $l$
as possible, the $l\rightarrow\infty$ limit being of interest. However,
the accuracy of TCSA decreases as $l$ grows. The accuracy can be
improved by taking higher $e_{cut}$, but this increases the size
of TCSA Hamiltonian and the time needed for diagonalization. Experience
shows that accuracy decreases for values $\beta$ near to $\sqrt{4\pi}$
(this is the value where UV divergences appear in conformal perturbation
theory) although the speed of convergence of the spectrum to the $l\rightarrow\infty$
asymptotic values increases, and the speed of convergence becomes
very low for values of $\beta$ near to $0$. Taking these properties
of the TCSA into consideration and following \cite{BPTW} we performed
calculations at the values $\beta=8\sqrt{\pi}/7$, $4\sqrt{\pi}/3$,
$8\sqrt{\pi}/5$. We also note that the accuracy of the TCSA spectra
is severely decreased if $V$ has several (local) minima.

\begin{figure}
\begin{center}\includegraphics[%
  width=0.17\paperwidth,
  height=0.40\paperwidth,
  angle=270]{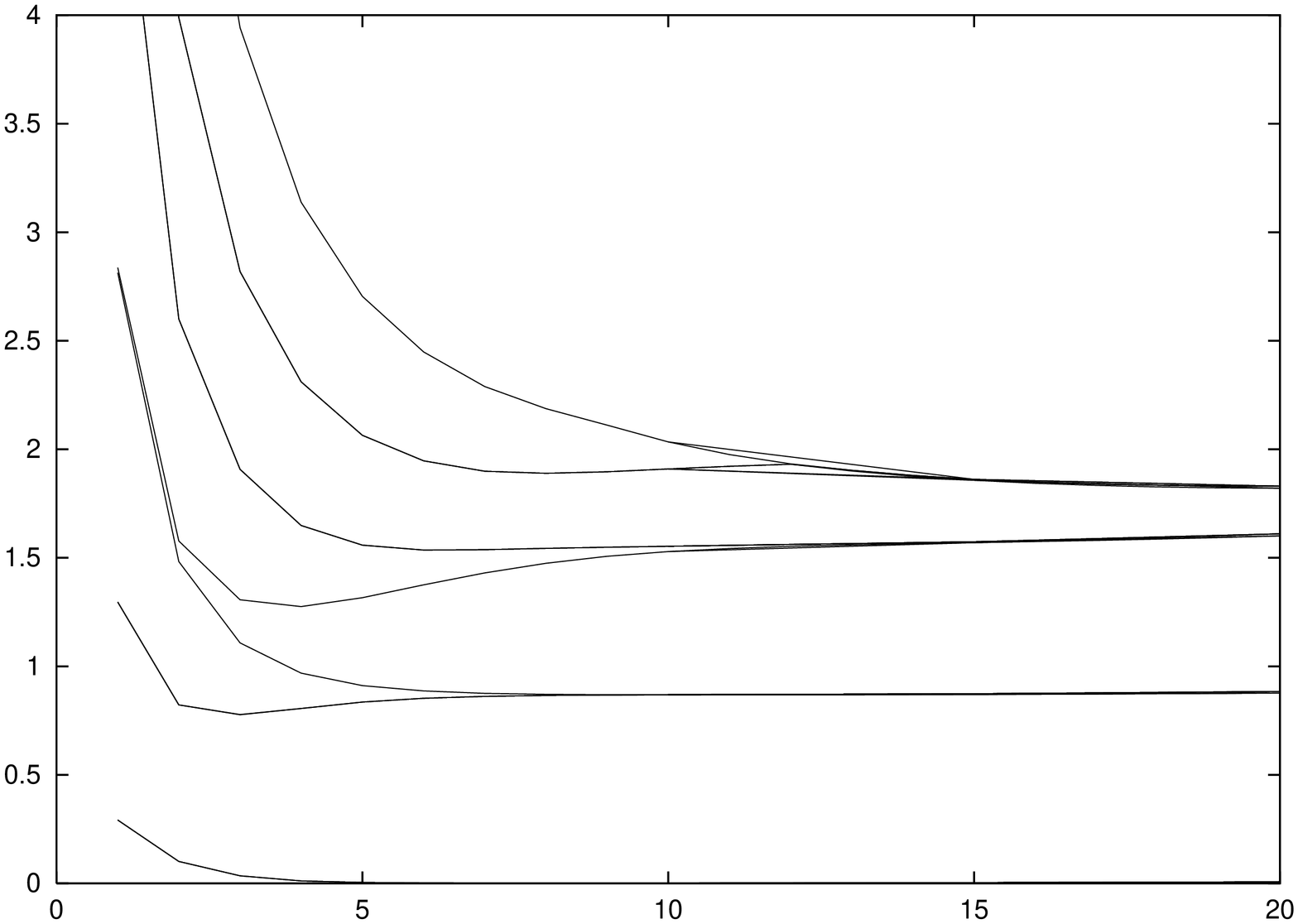}\includegraphics[%
  width=0.17\paperwidth,
  height=0.40\paperwidth,
  angle=270]{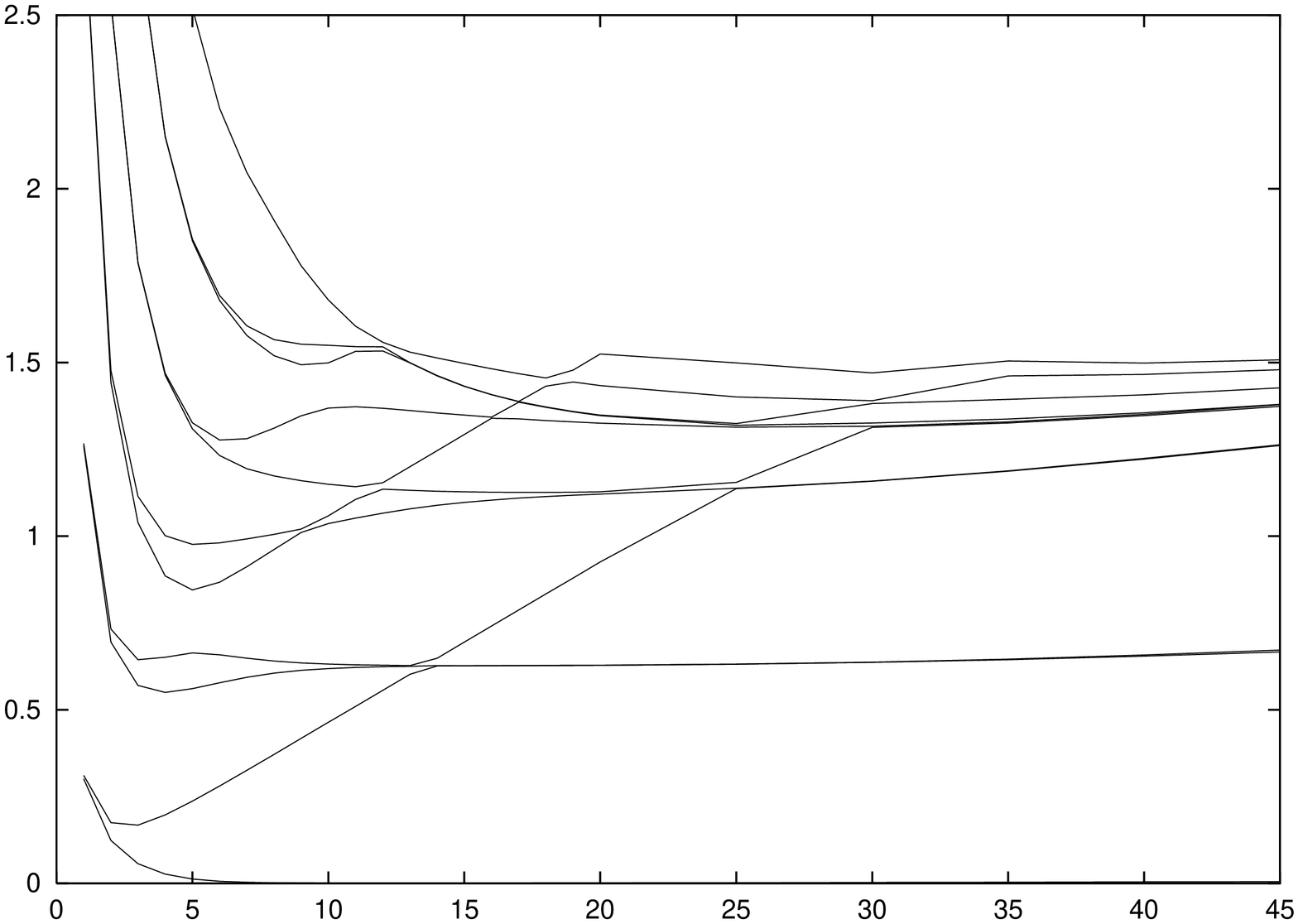}\end{center}

\vspace{-0.4cm} \hspace{3.5cm}a: $\tilde{\eta}=0$\hspace{7cm}b:
$\tilde{\eta}=0.3$ \vspace{-1cm}

\begin{center}\includegraphics[%
  width=0.17\paperwidth,
  height=0.40\paperwidth,
  angle=270]{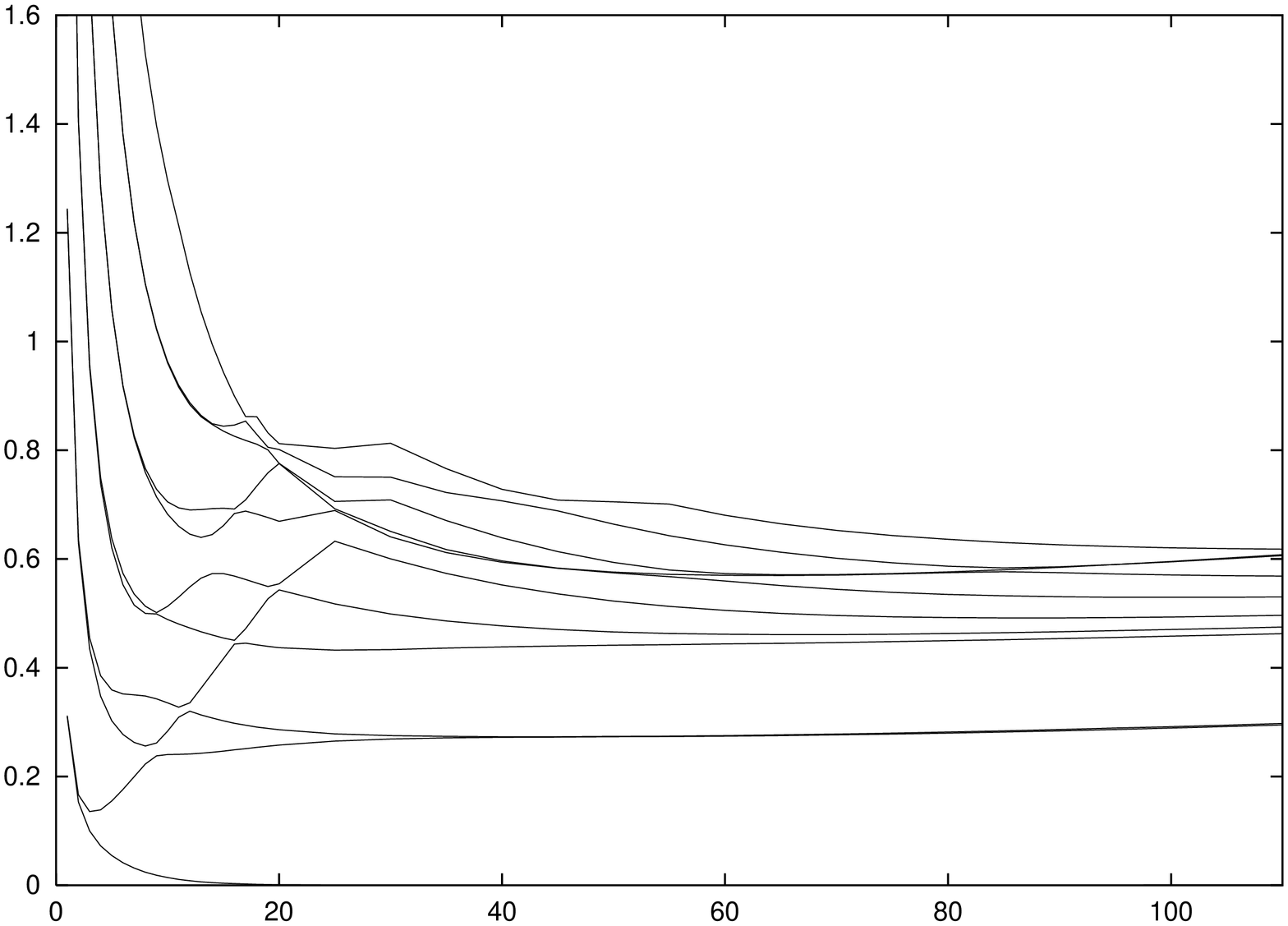}\includegraphics[%
  width=0.17\paperwidth,
  height=0.40\paperwidth,
  angle=270]{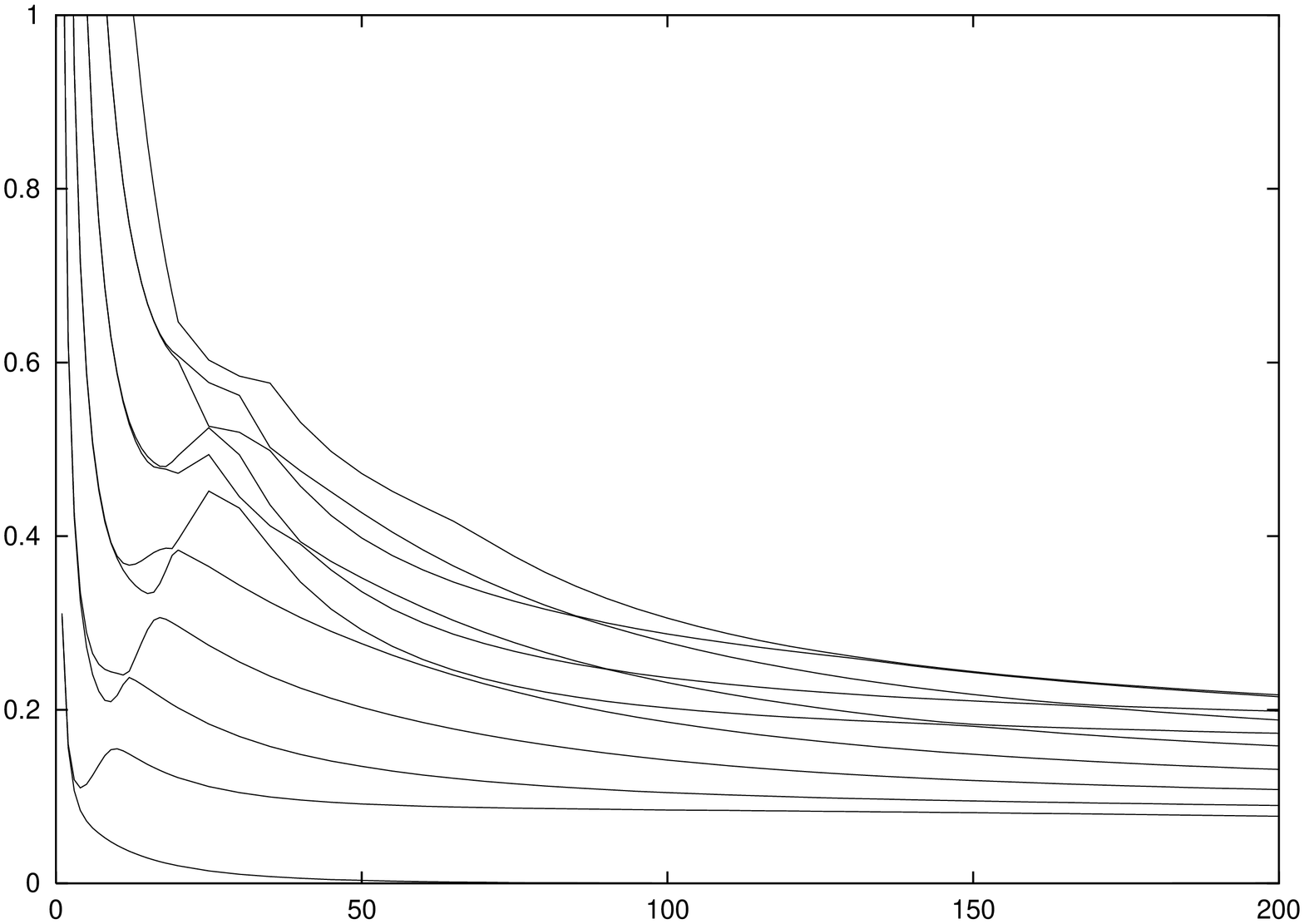}\end{center}

\vspace{-0.4cm} \hspace{3.5cm}c: $\tilde{\eta}=0.7$\hspace{6.5cm}d:
$\tilde{\eta}=0.92$ \vspace{-1cm}

\begin{center}\includegraphics[%
  width=0.17\paperwidth,
  height=0.40\paperwidth,
  angle=270]{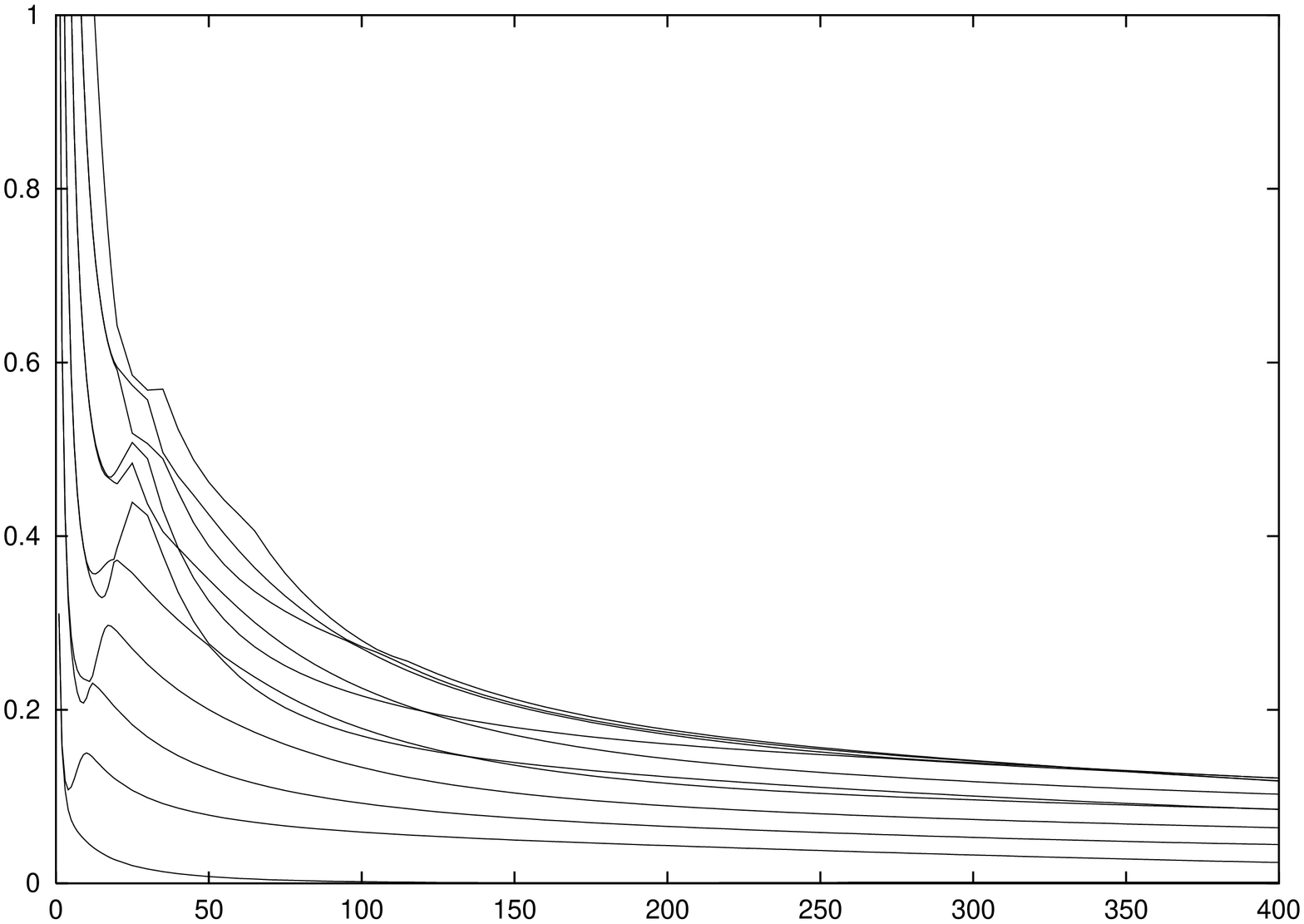}\includegraphics[%
  width=0.17\paperwidth,
  height=0.40\paperwidth,
  angle=270]{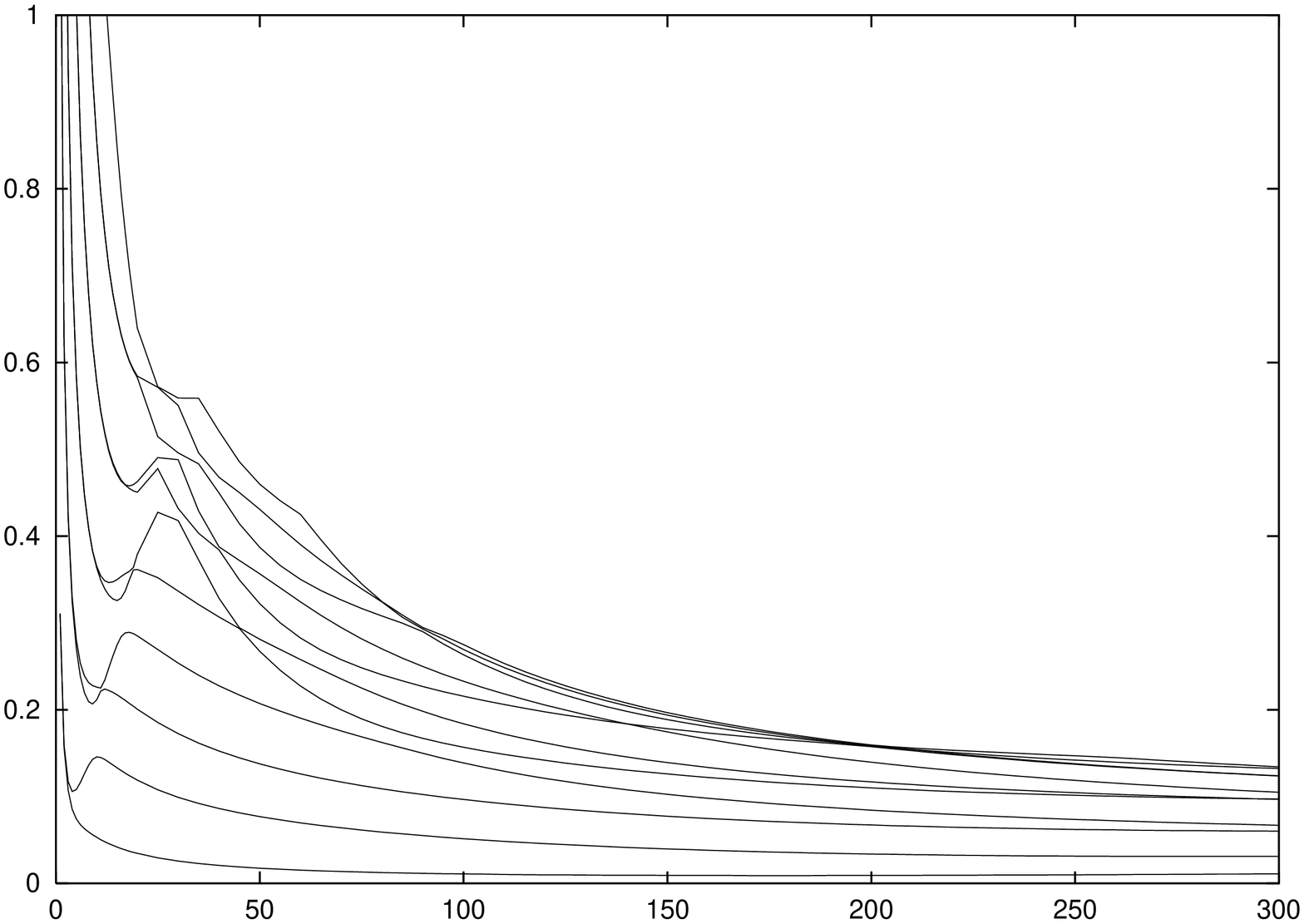}\end{center}

\vspace{-0.4cm} \hspace{3.5cm}e: $\tilde{\eta}=0.94$\hspace{6.3cm}f:
$\tilde{\eta}=0.96$ \vspace{-0.5cm}

\begin{center}\includegraphics[%
  width=0.17\paperwidth,
  height=0.40\paperwidth,
  angle=270]{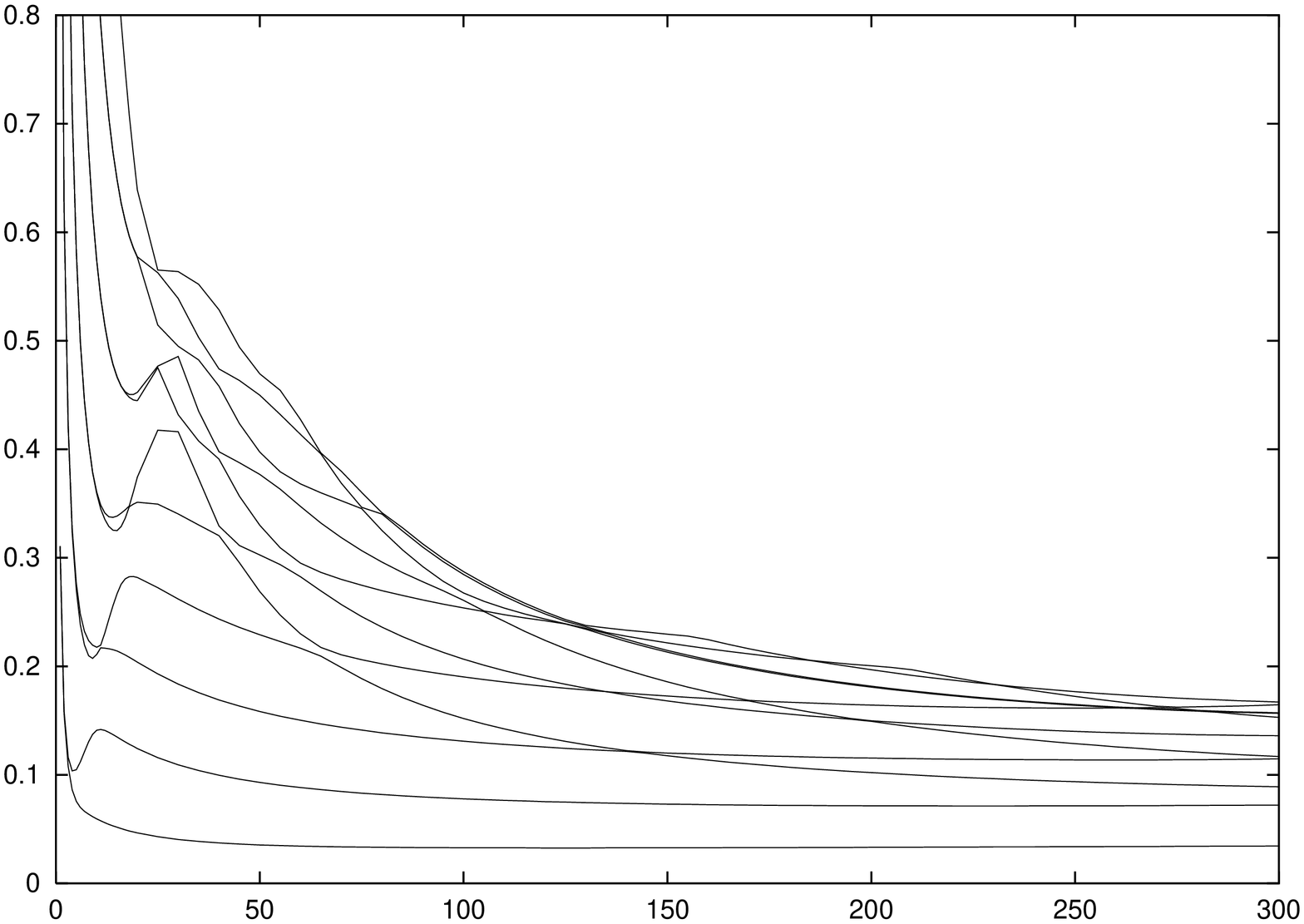}\end{center}

\begin{center}g: $\tilde{\eta}=0.98$\end{center}

\caption{\label{spectra1}Change of the spectrum as $\tilde{\eta}$ varies
from 0 to 1 at $\beta=4\sqrt{\pi}/3$. The first 12 energy levels
(including the ground level) relative to the ground level are shown
as functions of $l$. }
\end{figure}

Figure \ref{pot12}.a and \ref{pot12}.b show the shape of the classical
potential in the phases with broken and unbroken $\ZZ_{2}$-symmetry,
respectively. Figures \ref{spectra1}.a-\ref{spectra1}.g show TCSA
spectra obtained at $\beta=4\sqrt{\pi}/3$ at various values of $\tilde{\eta}$.
The TCSA Hilbert space had dimension $3700$, the first $12$ energy
levels are shown in the figures. The highest values of $l$ are chosen
so that the truncation error still be small (the massgap remain constant).
However, the effect of truncation is perceptible in Figure \ref{spectra1}.b
for instance. It can be seen that in the domain $\tilde{\eta}<0.92$
the ground states and the first massive states are doubly degenerate.
(They are triply degenerate at $\tilde{\eta}=0$.) {}``Runaway''
energy levels (of constant slope) corresponding to the single local
minimum of the potential can also be seen (especially clearly in Figure
\ref{spectra1}.b). In the domain $\tilde{\eta}>0.98$ the spectra
are massive, but the ground state and the first massive state
are nondegenerate. In the intermediate domain (especially for $\tilde{\eta}\sim0.95$)
the structure of the spectrum changes, no massgap and degeneracy can
clearly be seen. We obtained similar spectra at $\beta=8\sqrt{\pi}/5$
and $\beta=8\sqrt{\pi}/7$ as well. As we did not see {}``runaway''
energy levels that would have signaled first order phase transition
in the transitional domain of $\tilde{\eta}$ we analyzed the data
by looking for a second order Ising type phase transition at some
critical values $\tilde{\eta}_{c}(\beta)$.

The Ising model contains three primary fields: the identity with weights
$(0,0)$, the $\epsilon$ with weights $(1/2,1/2)$, and $\sigma$
with weights $(1/16,1/16)$. Since the DSG model exhibits the $\ZZ_{2}$-symmetry
for all values of $\tilde{\eta}$, the $\ZZ_{2}$-odd $\sigma$ and
its descendants cannot appear as perturbations in the Hamiltonian
(\ref{IR Hamilton}). The only relevant field compatible with the
$\ZZ_{2}$-symmetry is $\epsilon$ (the contribution of the identity
cancels in the relative energy levels). The presence of a relevant
perturbation $\epsilon$ in the Hamiltonian leads to a correction
$B_{\Psi}$ or $B_{\Psi}+C_{\Psi}l$ to $e_{\Psi}(l)-e_{0}(l)$. The
term $C_{\Psi}l$ is of second order in perturbed CFT. The leading
irrelevant perturbation (compatible with the $\ZZ_{2}$-symmetry)
is the first descendant of $\epsilon$, this gives a correction $A_{\Psi}l^{-2}$
to $e_{\Psi}(l)-e_{0}(l)$ (in first order). Thus we expect that in
a large but finite volume range, near $\tilde{\eta}_{c}(\beta)$,
the volume dependence of the energy levels is described well by the
formula\begin{equation}
e_{i}(l)-e_{0}(l)=\frac{2\pi}{l}D_{i}+A_{i}l^{-2}+B_{i}+C_{i}l.\label{fit1}\end{equation}
 We fitted this function to the lowest energy levels obtained by TCSA
and determined the {}``best'' $\tilde{\eta}_{c}(\beta)$ value by
tuning $\tilde{\eta}$ in the transition region and looking for whether
$e_{2}(l)-e_{0}(l)$ continues to decrease along the complete $l$
range ($\lim_{l\rightarrow\infty}(e_{2}(l)-e_{0}(l))=0$ only at $\tilde{\eta}_{c}(\beta)$),
and $B_{i}$ and $C_{i}$ are as small as possible. The result is
shown in Table \ref{tabla1}. The fitting was done in the volume ranges
$l=10-105$, $l=55-105$; $l=10-140$, $l=100-190$; $l=20-200$,
$l=150-390$. The errors presented come from the fitting process and
do not contain the truncation errors which are generally much larger.%

\noindent The first two energy levels above the ground state correspond
to the operators $\sigma$ and $\epsilon$ in the Ising model. These
operators have conformal weights $\Delta^{\pm}=1/16$ and $\Delta^{\pm}=1/2$,
so the exact values for $D_{1}$ and $D_{2}$ are \[
D_{1}=0.125,\qquad D_{2}=1.\]
 The results of the fits agree quite well with this prediction.

\begin{figure}
\begin{center}\includegraphics[%
  width=0.90\columnwidth,
  height=0.27\textwidth]{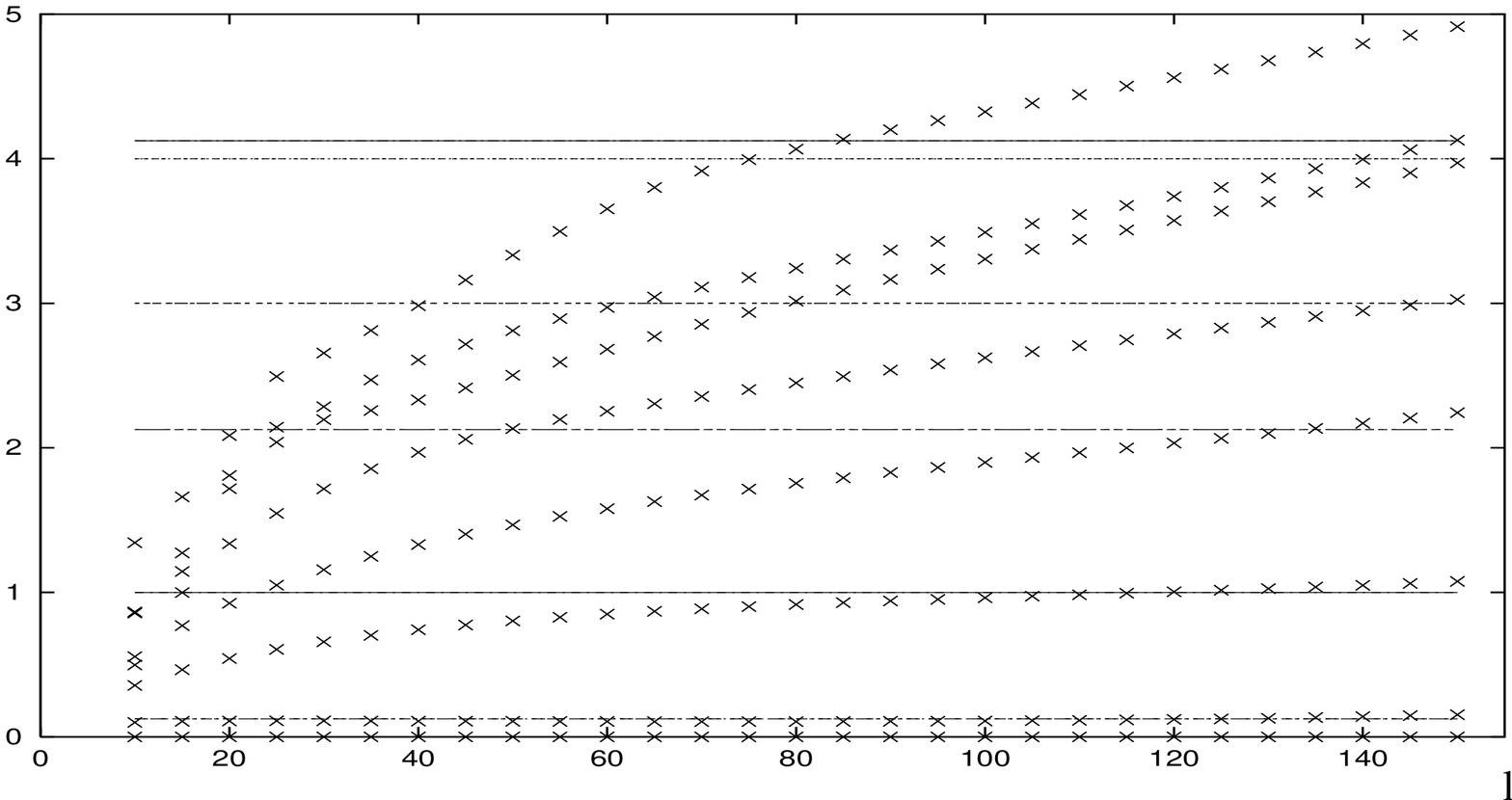}\end{center}

\begin{center}$[e_{i}(l)-e_{0}(l)]\cdot l/2\pi$ at $\beta=8\sqrt{\pi}/5$
and $\tilde{\eta}=0.944$\end{center}

\begin{center}\includegraphics[%
  width=0.90\columnwidth,
  height=0.27\textwidth]{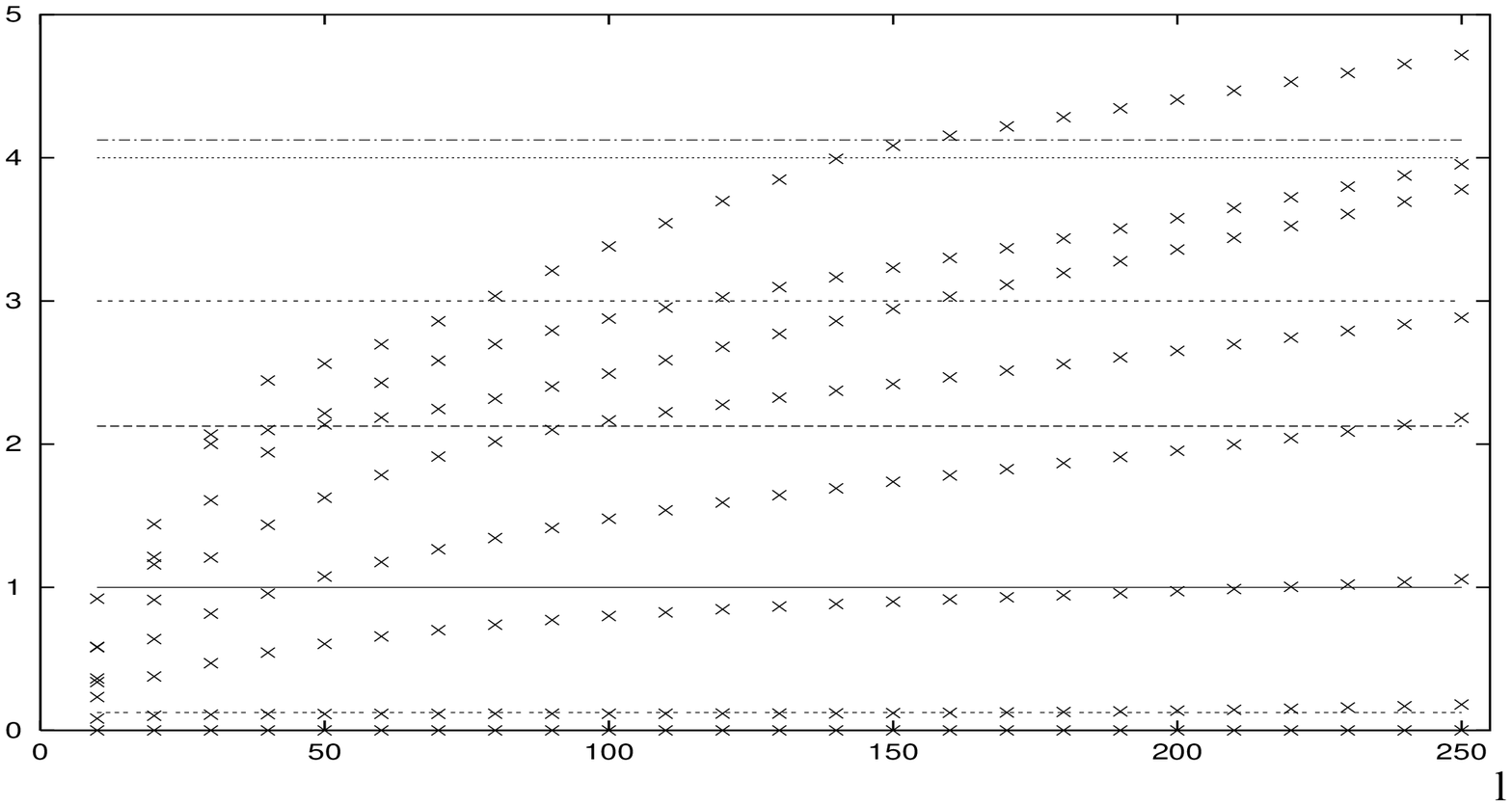}\end{center}

\begin{center}$[e_{i}(l)-e_{0}(l)]\cdot l/2\pi$ at $\beta=4\sqrt{\pi}/3$
and $\tilde{\eta}=0.955$\end{center}

\begin{center}\includegraphics[%
  width=0.90\columnwidth,
  height=0.27\textwidth]{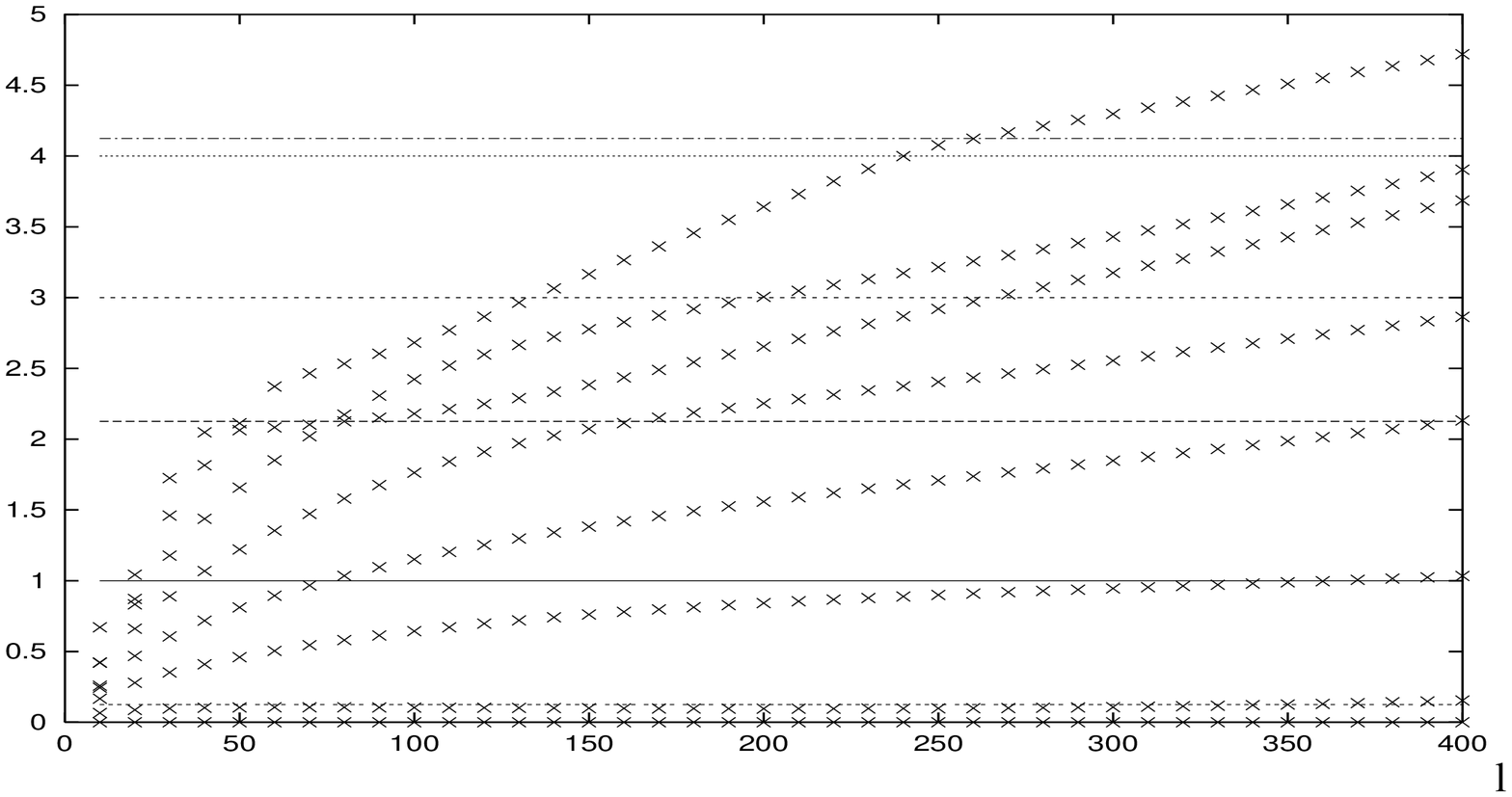}\end{center}

\begin{center}$[e_{i}(l)-e_{0}(l)]\cdot l/2\pi$ at $\beta=8\sqrt{\pi}/7$
and $\tilde{\eta}=0.961$\end{center}

\caption{\label{critspectra}TCSA spectra as functions of $l$ at the estimated
critical values of $\tilde{\eta}$ }
\end{figure}

The TCSA data obtained at the estimated values of $\tilde{\eta}_{c}$
using a truncated space with dimension 4800, 5300 and 5100 are shown
in Figure \ref{critspectra}. These figures show energy levels multiplied
by $l/(2\pi)$ as functions of $l$. The constant lines corresponding
to the Ising model values of $D_{i}$ are also shown in these figures.
\begin{table}
\begin{center}\begin{tabular}{|c|c|c|c|c|}
\hline 
State&
 $D_{i}$&
 $A_{i}$&
 $B_{i}$&
 $C_{i}$\tabularnewline
\hline
$i=1$&
 $0.138\pm0.0005$&
 $-2.0\pm0.02$&
 $-0.0046\pm0.0001$&
 $2.98\cdot10^{-5}\pm9\cdot10^{-7}$\tabularnewline
\hline
$i=2$&
 $1.00\pm0.01$&
 $-74\pm2$&
 $0.004\pm0.001$&
 $9\cdot10^{-6}\pm4\cdot10^{-6}$ \tabularnewline
\hline
\end{tabular}\end{center}

\begin{center}$\beta=8\sqrt{\pi}/5$, $\tilde{\eta}=0.944$\end{center}

\begin{center}\begin{tabular}{|c|c|c|c|c|}
\hline 
State&
 $D_{i}$&
 $A_{i}$&
 $B_{i}$&
 $C_{i}$\tabularnewline
\hline
$i=1$&
 $0.125\pm0.001$&
 $-2.63\pm0.025$&
 $-0.0002\pm0.0001$&
 $3\cdot10^{-7}\pm1\cdot10^{-6}$\tabularnewline
\hline
$i=2$&
 $1.04\pm0.02$&
 $-152\pm6$&
 $0.0014\pm0.0009$&
 $-1.5\cdot10^{-5}\pm2\cdot10^{-6}$ \tabularnewline
\hline
\end{tabular}\end{center}

\begin{center}$\beta=4\sqrt{\pi}/3$, $\tilde{\eta}=0.955$\end{center}

\begin{center}\begin{tabular}{|c|c|c|c|c|}
\hline 
State&
 $D_{i}$&
 $A_{i}$&
 $B_{i}$&
 $C_{i}$\tabularnewline
\hline
$i=1$&
 $0.125\pm0.001$&
 $-4.4\pm0.1$&
 $-0.0009\pm0.0001$&
 $4\cdot10^{-7}\pm5\cdot10^{-7}$\tabularnewline
\hline
$i=2$&
 $1.00\pm0.02$&
 $-252\pm9$&
 $0.0002\pm0.0004$&
 $4.3\cdot10^{-6}\pm6\cdot10^{-7}$ \tabularnewline
\hline
\end{tabular}\end{center}

\begin{center}$\beta=8\sqrt{\pi}/7$, $\tilde{\eta}=0.961$\end{center}

\vspace*{\medskipamount}

\caption{\label{tabla1}The results of fitting (\ref{fit1}) to the first
two excited levels for various values of $\beta$ at the estimated
critical value of $\tilde{\eta}$ }
\end{table}

To summarize, evaluating the TCSA data we found that the phase transition
is second order and Ising type at the values of $\beta$ chosen. This
statement is true up to the precision of the TCSA data, which is about
$10^{-2}-10^{-3}$ for the values of the volume used. We remark that
it is not possible to distinguish a second order transition from a
(very) weakly first order transition by TCSA. %

For values of $\beta$ near $0$ we expect second order phase transition,
because the model is semi-classical in this region and so the correction
to the classical potential in the effective potential is expected
to be small.

A correction to the classical potential in the effective potential
of frequency $2\beta/3$ is possible in principle. A correction with
this frequency is ---unlike in the case of $\alpha/\beta=1/2$---
relevant for any values of $\beta$, and it can be verified by elementary
calculation that it may change the order of the transition outside
the semiclassical region if its coefficient is sufficiently large
(see also \cite{FGN,BPTW}). However, the accuracy of TCSA did not
allow us to perform calculations at values about $\beta^{2}>8\pi/3$
and to check the nature of phase transition in this domain.

\begin{figure}
\begin{center}\includegraphics[%
  width=0.85\textwidth]{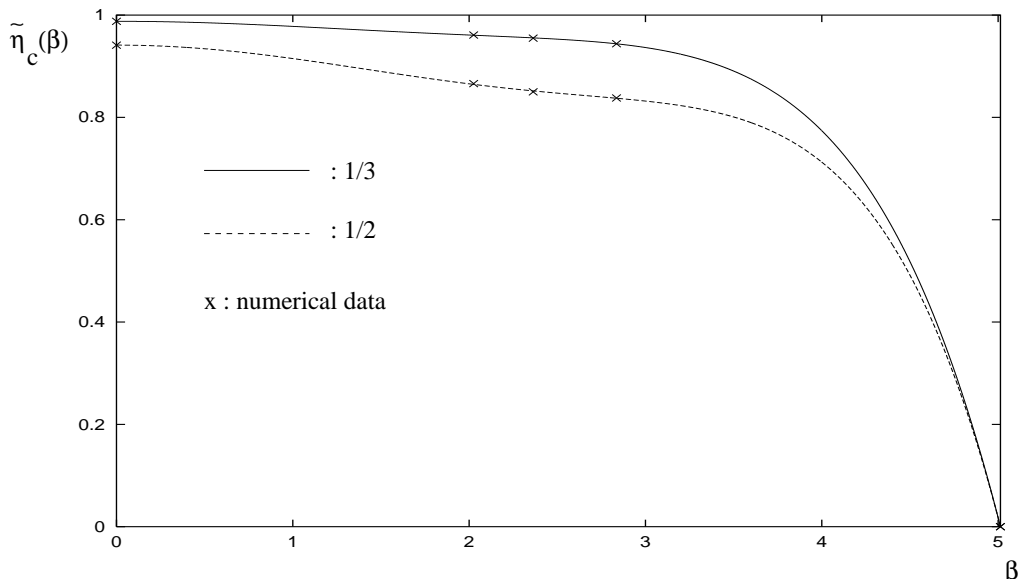}\end{center}

\caption{\label{phdiag1}The phase diagram of the two-frequency model at $\alpha/\beta=1/2$ and $\alpha/\beta=1/3$}
\end{figure}

The phase diagram in the $(\beta,\tilde{\eta})$ plane based on the
data obtained by TCSA can be seen in Figure \ref{phdiag1}. We took
into consideration that $\tilde{\eta}_{c}(0)=3^{4}/(1+3^{4})\approx0.988$
is exactly known $\beta=0$ being the classical limit, and at $\beta=\sqrt{8\pi}$
the term with frequency $\beta$ in the potential becomes irrelevant
and thus for $\beta\rightarrow\sqrt{8\pi}$ the other term and so
the symmetric phase is expected to dominate, so $\lim_{\beta\rightarrow\sqrt{8\pi}}\tilde{\eta}_{c}(\beta)=0$.
The figure shows the three values of $\tilde{\eta}_{c}$ obtained
from the TCSA data and the two values at $\beta=0$ and $\beta=\sqrt{8\pi}$.
The continuous line is obtained by fitting an even polynomial (note
that $\tilde{\eta}_{c}(\beta)=\tilde{\eta}_{c}(-\beta)$) to these
values and is shown in order to get an idea of the phase transition
line. The model is in the symmetric phase above the line and in the
phase with broken symmetry below the line. The data obtained by \cite{BPTW}
for the $\alpha/\beta=1/2$ case are also shown, the dashed line is
fitted to these data.

\section{\label{sec: het}Phase diagram of the three-frequency model}

We take the same values of the parameters of (\ref{eq: 3freq pot})
as in section \ref{sec 4.2}, namely $\beta_{1}=\beta,\beta_{2}=\frac{2}{3}\beta,\beta_{3}=\frac{1}{3}\beta,$
$\delta_{2}=\delta_{3}=0$, $\mu_{1},\mu_{3}<0$, $\mu_{2}>0$, and
investigate the quantum model in this case.

\subsection{The tricritical point}

\begin{figure}
\begin{center}\includegraphics[%
  height=0.70\paperwidth,
  angle=270]{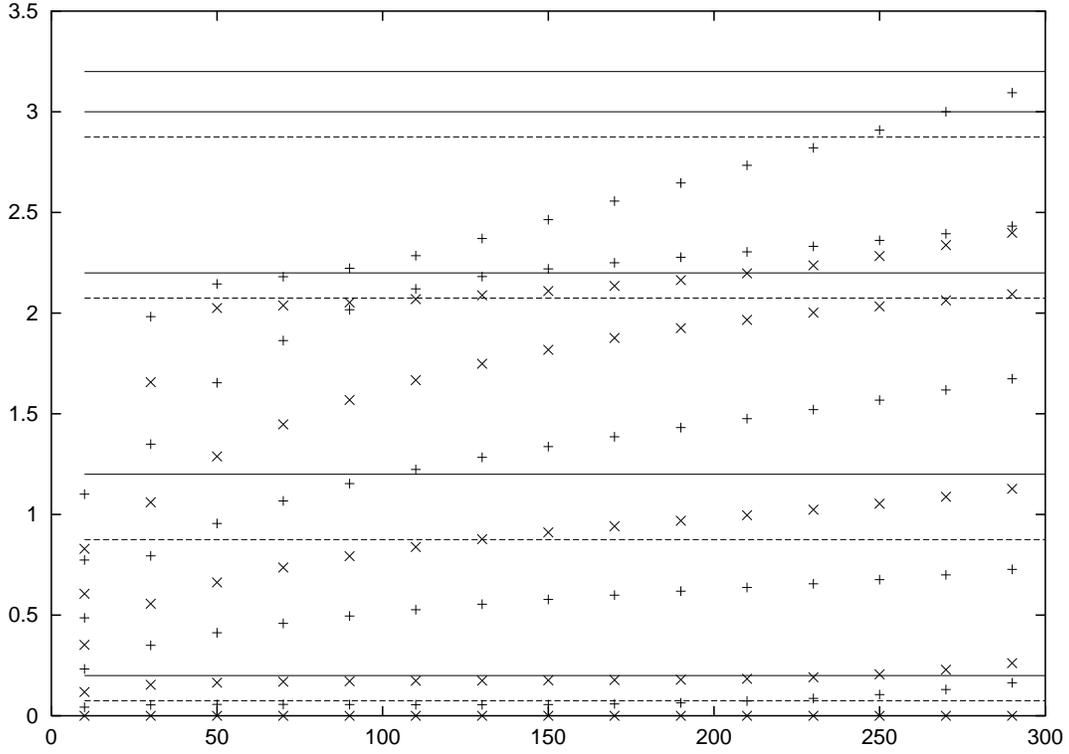}\end{center}

\caption{\label{spectrum3}$[e_{i}(l)-e_{0}(l)]\cdot l/2\pi$ as functions
of $l$ obtained by TCSA at $\beta=8\sqrt{\pi}/7$, $\eta_{1}=0.163$,
$\eta_{2}=0.3518$ }
\end{figure}

The tricritical Ising model contains 6 primary fields with the following
conformal weights:\begin{equation}
(0,0),\left(\frac{1}{10},\frac{1}{10}\right),\left(\frac{3}{5},\frac{3}{5}\right),\left(\frac{3}{2},\frac{3}{2}\right),\label{eq:tri1}\end{equation}
 and\begin{equation}
\left(\frac{3}{80},\frac{3}{80}\right),\left(\frac{7}{16},\frac{7}{16}\right).\label{eq:tri2}\end{equation}
 The fields corresponding to (\ref{eq:tri1}) are even and those corresponding
to (\ref{eq:tri2}) are odd with respect to the parity, so only 
the fields corresponding to (\ref{eq:tri1}) and their descendants
can contribute to the Hamiltonian (\ref{IR Hamilton}). Thus the volume
dependence of the energy levels near the tricritical point should
be described well for large $l$ by \[
e_{\Psi}(l)-e_{0}(l)=\frac{2\pi}{l}(\Delta_{IR,\Psi}^{+}+\Delta_{IR,\Psi})+A_{\Psi}l^{-0.2}+B_{\Psi}l^{0.8}+...,\]
 where only the leading terms are kept. Searching for the tricritical
point we fitted the function \begin{equation}
\frac{2\pi}{l}D_{i}+A_{i}l^{-0.2}+B_{i}l^{0.8}\label{eq:fit2}\end{equation}
 to the data obtained by TCSA for $e_{i}(l)-e_{0}(l)$ near the estimated
location of the tricritical point. Best fits are shown in Table \ref{table2}.
(The errors presented come from the fitting process and do not contain
the truncation errors which are generally much larger.) The fitting
was done in the volume ranges $l=50-230$, $l=110-230$, the dimension
of the truncated Hilbert space was 13600. The results of the fitting
support the existence of a tricritical point located (approximately)
at $\eta_{1}=0.163$, $\eta_{2}=0.3518$. The exact values of $D_{1}$
and $D_{2}$ in the tricritical Ising model are \[
D_{1}=0.075,\qquad D_{2}=0.2.\]
The numerical results agree quite well with this prediction. The
TCSA spectrum obtained at the tricritical point is shown in Figure
\ref{spectrum3}. The values of $D_{i}$ predicted by the tricritical
Ising model are also shown in the figure. The dashed lines and $+$
signs are used for odd parity states, the continuous lines and $\times$
signs are used for even parity states. %
\begin{table}
\begin{center}\begin{tabular}{|c|c|c|c|}
\hline 
State&
 $D_{i}$&
 $A_{i}$&
 $B_{i}$\tabularnewline
\hline
$i=1$&
 $0.074\pm0.004$&
 $-0.0060\pm0.001$&
 $2.8\cdot10^{-5}\pm5\cdot10^{-6}$\tabularnewline
\hline
$i=2$&
 $0.196\pm0.01$&
 $-0.006\pm0.002$&
 $2.4\cdot10^{-5}\pm6\cdot10^{-6}$ \tabularnewline
\hline
\end{tabular}\end{center}

\vspace*{\medskipamount}

\begin{center}$\beta=8\sqrt{\pi}/7$, $\eta_{1}=0.163$, $\eta_{2}=0.3518$\end{center}

\vspace*{\medskipamount}

\caption{\label{table2}The results of fitting (\ref{eq:fit2}) to the first
two excited levels in the estimated tricritical point }
\end{table}

We used a modified version of the TCSA code exploiting the $\ZZ_{2}$-symmetry
of the model by taking even and odd basis vectors and taking the 
Hamiltonian on the even and odd subspaces
separately, which reduces the total time needed for diagonalization
and thus allows to take higher $e_{cut}$ values.

The TCSA data for the first two excited states fit quite well to the
prediction of the tricritical Ising model, the energy levels of the
next two excited states also show correspondence with the prediction.
Clear correspondence cannot be seen for higher levels.

\subsection{The critical line}

\begin{table}
\begin{center}\begin{tabular}{|c|c|c|c|c|c|c|c|}
\hline 
$\eta_{1}$&
 $\eta_{2}$&
 $\eta_{1}$&
 $\eta_{2}$&
 $\eta_{1}$&
 $\eta_{2}$&
 $\eta_{1}$&
 $\eta_{2}$\tabularnewline
\hline
0.01&
 0.354&
 0.06&
 0.355&
 0.11&
 0.359&
 0.16 &
 0.357\tabularnewline
\hline
0.02&
 0.354&
 0.07&
 0.3555&
 0.12&
 0.36&
 0.163&
 0.3565\tabularnewline
\hline
0.03&
 0.354&
 0.08&
 0.356&
 0.13&
 0.36&
&
\tabularnewline
\hline
0.04&
 0.354&
 0.09&
 0.3575&
 0.14&
 0.3595&
&
\tabularnewline
\hline
0.05&
 0.354&
 0.1&
 0.3585&
 0.15&
 0.359&
&
 \tabularnewline
\hline
\end{tabular}\end{center}

\caption{\label{critline}Points of the critical line found by TCSA}
\end{table}

\begin{figure}[H]
\begin{center}\includegraphics[%
  width=0.17\paperwidth,
  height=0.40\paperwidth,
  angle=270]{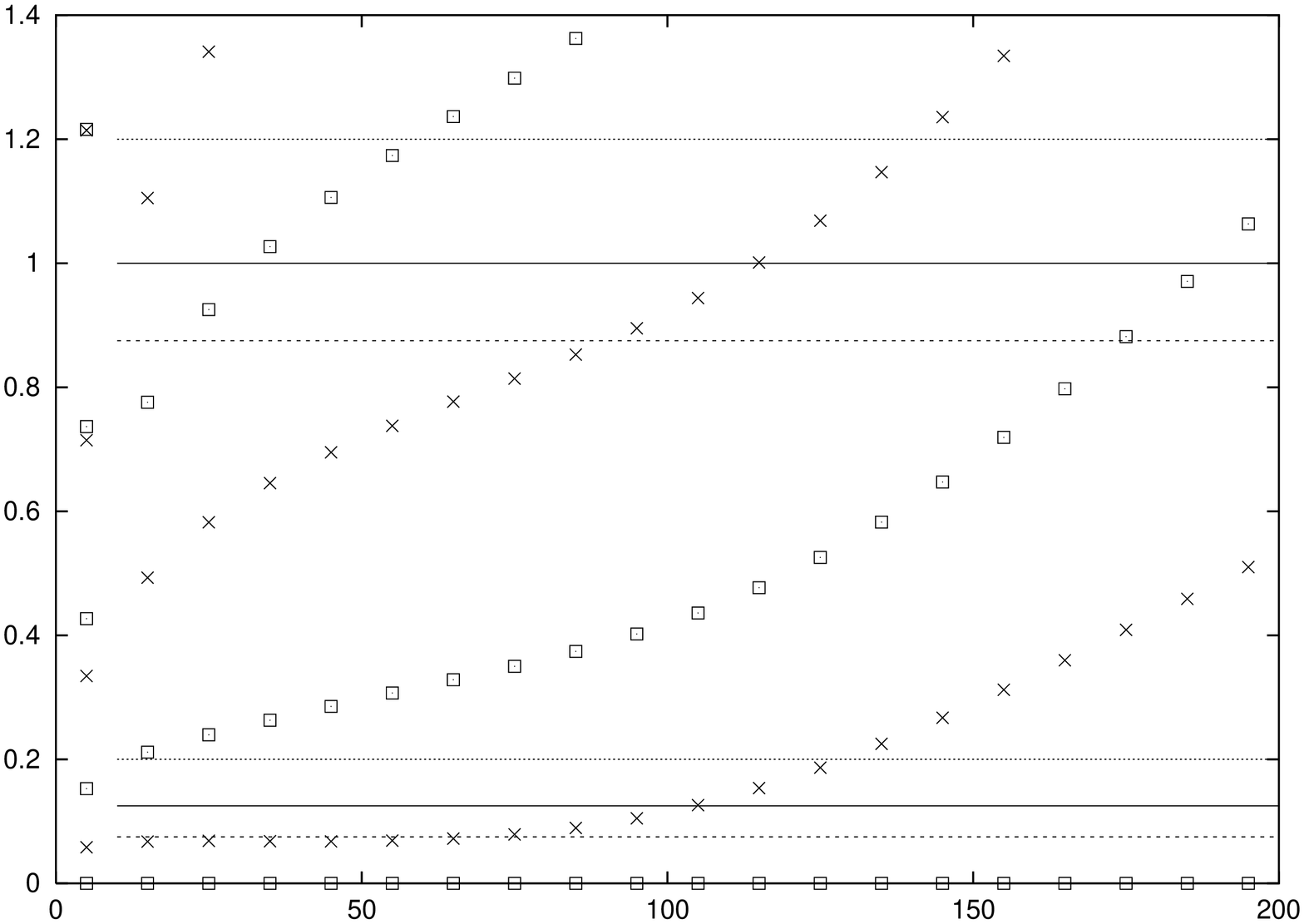}\includegraphics[%
  width=0.17\paperwidth,
  height=0.40\paperwidth,
  angle=270]{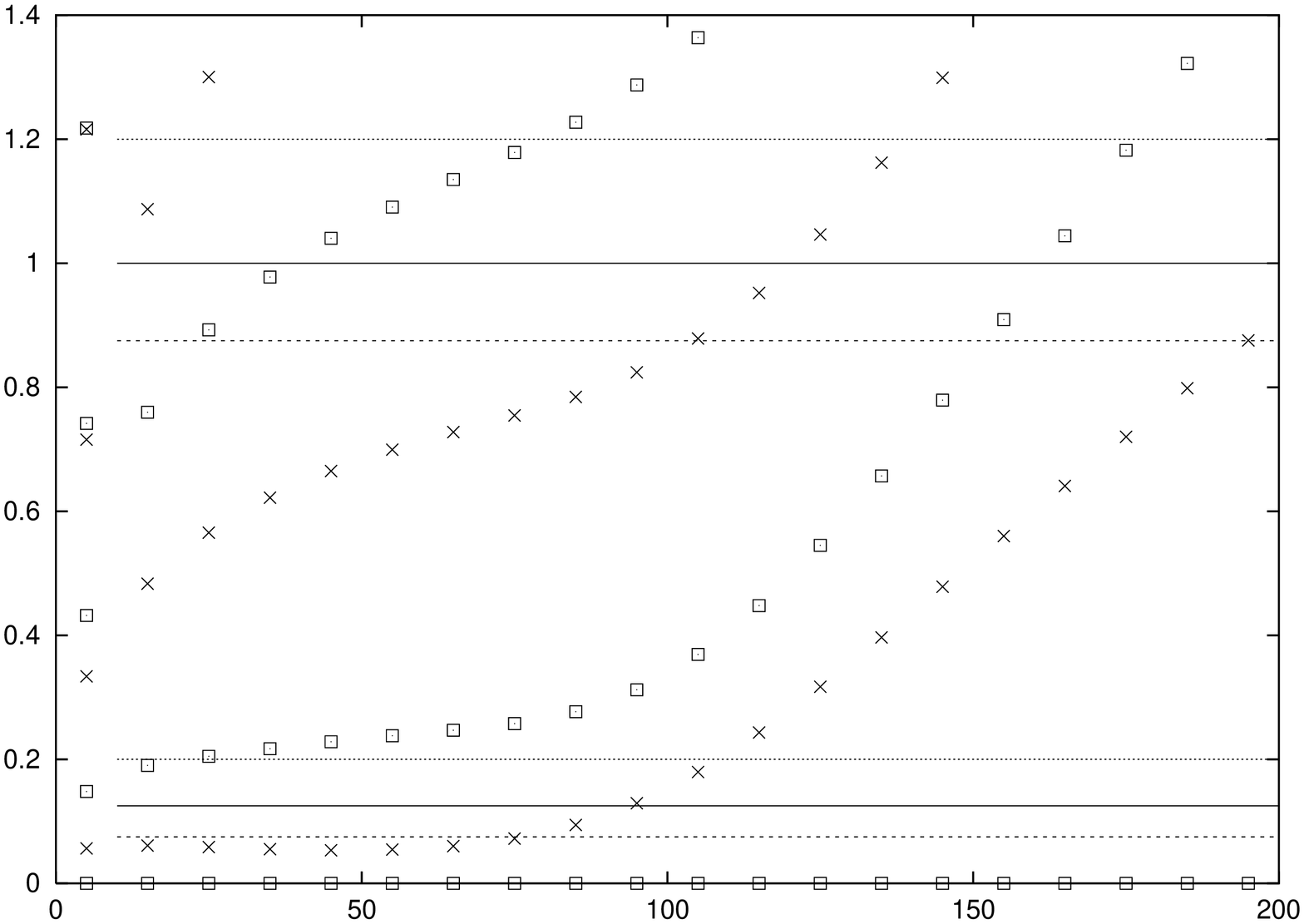}\end{center}

\vspace{-0.4cm} \hspace{3.5cm}i: $\eta_{1}=0.14$\hspace{6cm}j:
$\eta_{1}=0.15$ \vspace{-1cm}

\begin{center}\includegraphics[%
  width=0.17\paperwidth,
  height=0.40\paperwidth,
  angle=270]{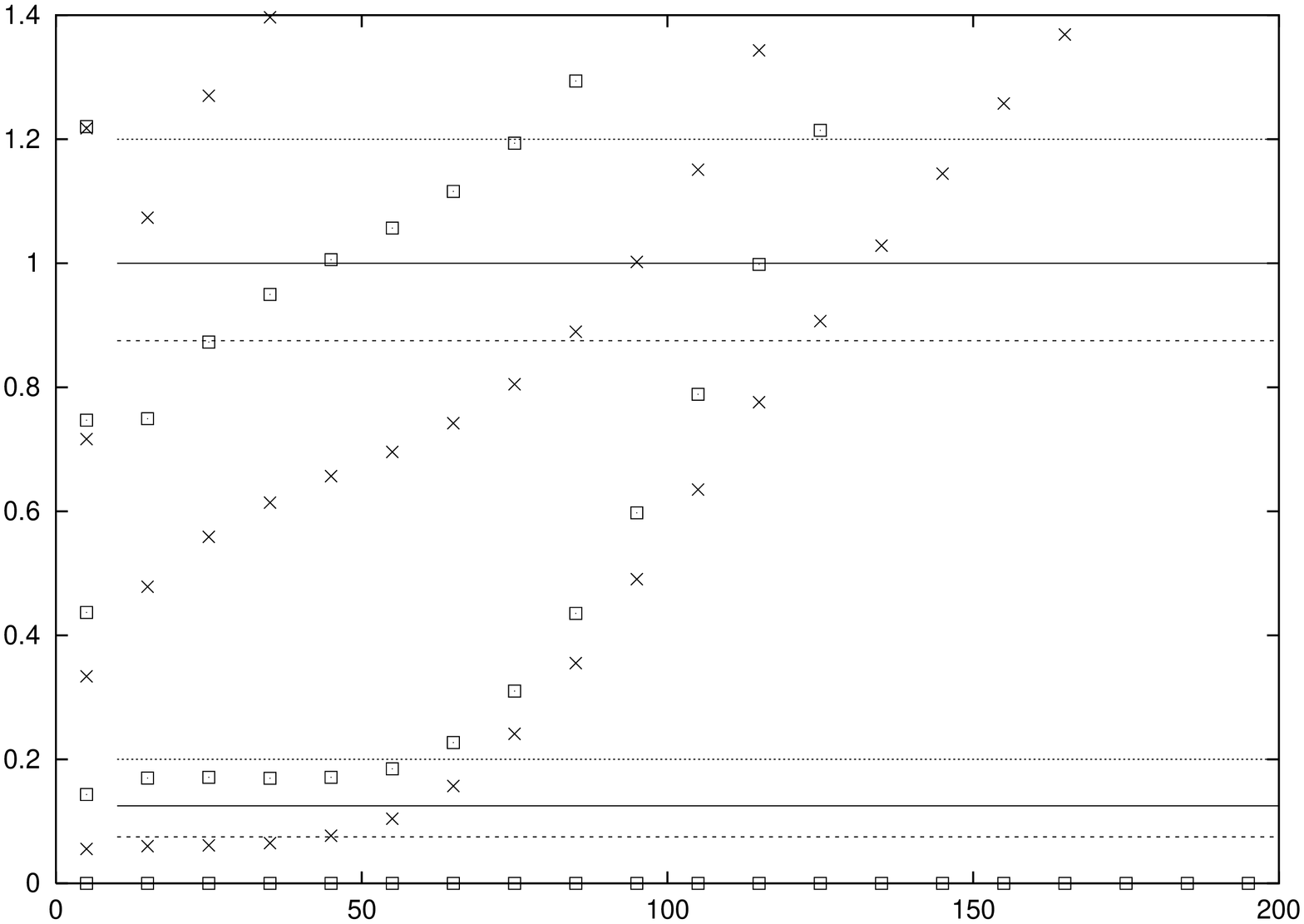}\includegraphics[%
  width=0.17\paperwidth,
  height=0.40\paperwidth,
  angle=270]{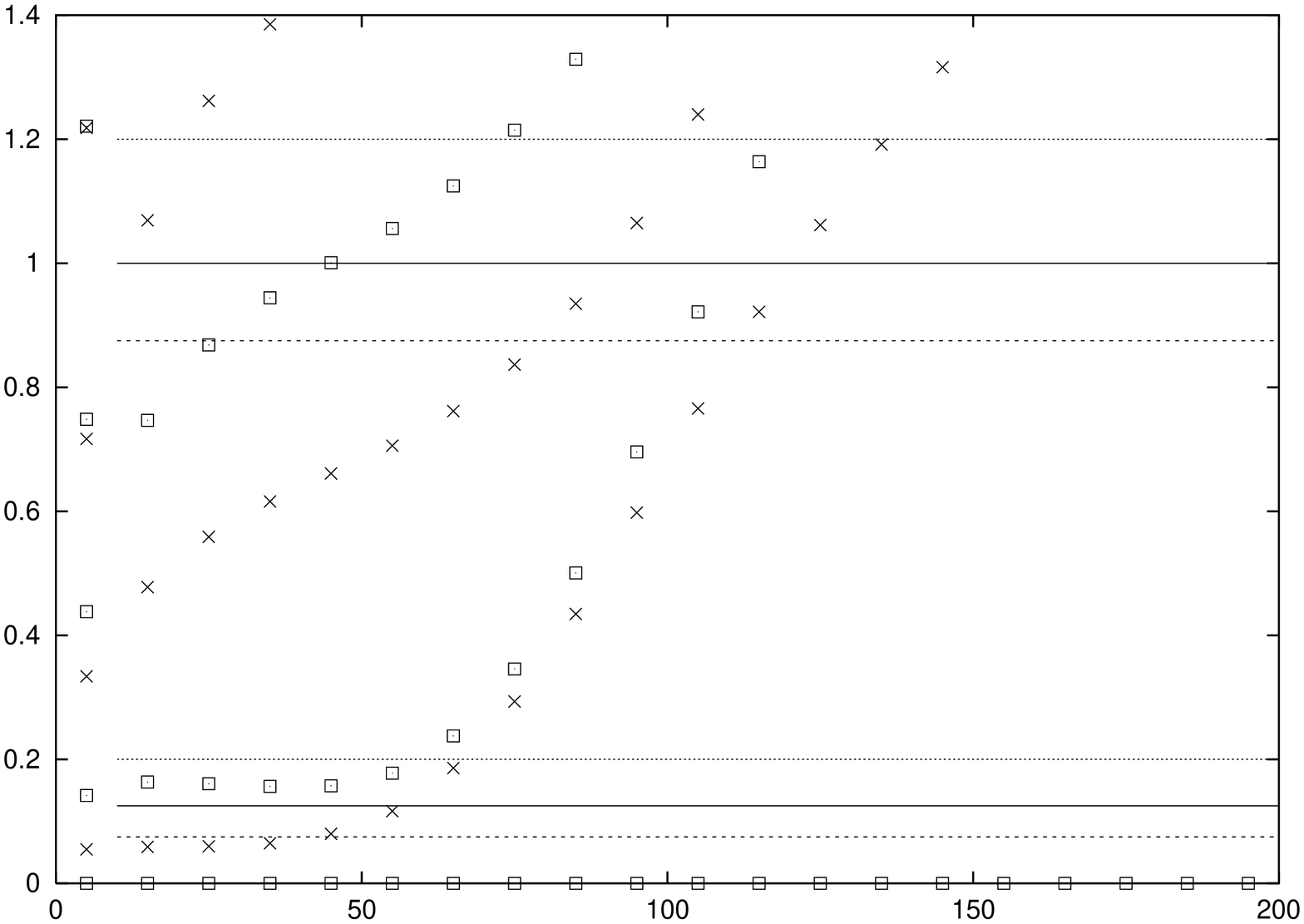}\end{center}

\vspace{-0.4cm} \hspace{3.5cm}k: $\eta_{1}=0.16$\hspace{6cm}l:
$\eta_{1}=0.163$

\caption{\label{dikrit vonal}$[e_{i}(l)-e_{0}(l)]\cdot l/2\pi$ as functions
of $l$ obtained by TCSA at $\beta=8\sqrt{\pi}/5$ and at various
points $(\eta_{1},\eta_{2})$ lying on the critical line, the predictions
of the critical Ising model (continuous horizontal lines) for $D_{i}$,
the predictions of the tricritical Ising model (dashed horizontal
lines) for $D_{i}$ }
\end{figure}

The points of the critical line we found using TCSA are listed in
Table \ref{critline}. The value of $\eta_{1}$ was chosen and fixed
in advance, and then $\eta_{2}$ was estimated in the same way as
described in Section \ref{sec: hat}. These points are also marked
in Figure \ref{threefreq phasediag} by crosses. The dimension of
the truncated Hilbert space was $10269$ in these calculations, which
corresponded to $e_{cut}=17$. The value of $\beta$ was $8\sqrt{\pi}/5$.
Figures \ref{dikrit vonal}.a-l show the TCSA spectra (especially
the lowest lying energy levels) obtained in these points as well as
the values of $D_{i}$ corresponding to both the critical and the
tricritical Ising model. Crosses are used for odd parity states, squares 
are used for even parity states.
It can be seen that moving on the critical
line in the phase space towards the tricritical endpoint the finite
volume spectrum changes continuously. In the $\eta_{1}<0.11$ domain
the spectra (especially the first two levels) correspond clearly to
phase transitions in the Ising universality class. At $\eta_{1}=0.11$
the first excited level already appears to correspond to the prediction
of the tricritical Ising model, whereas the second excited level still
has the behaviour predicted by the Ising model. It would be very interesting
to know the large volume behaviour (and infinite volume limit) of
the first excited level, but the precision of TCSA does not allow
to determine it. What we can see is that there is no sign in the TCSA
data that the first excited level follows the predictions of the Ising
model in the large volume limit. In the domain $0.11\leq\eta_{1}\leq0.16$
there is a spectacular rearrangement of the higher energy levels (already
observable in the $\eta_{1}<0.11$ domain), and at $\eta_{1}=0.16$
the second excited level also appears to correspond to the prediction
of the tricritical Ising model. We regard therefore the point $\eta_{1}=0.16$,
$\eta_{2}=0.357$ to be the tricritical endpoint of the critical line,
this point is marked by a square in Figure \ref{threefreq phasediag}.
(We did not aspire to determine the value of $\eta_{1}$ more precisely
for this value of $\beta$.) Each figure shows the spectrum in the
volume interval $l=0..200$, and for the large values $l\approx200$
the truncation error is always conspicuous.%
\begin{figure}[H]
\begin{center}\includegraphics[%
  width=0.17\paperwidth,
  height=0.40\paperwidth,
  angle=270]{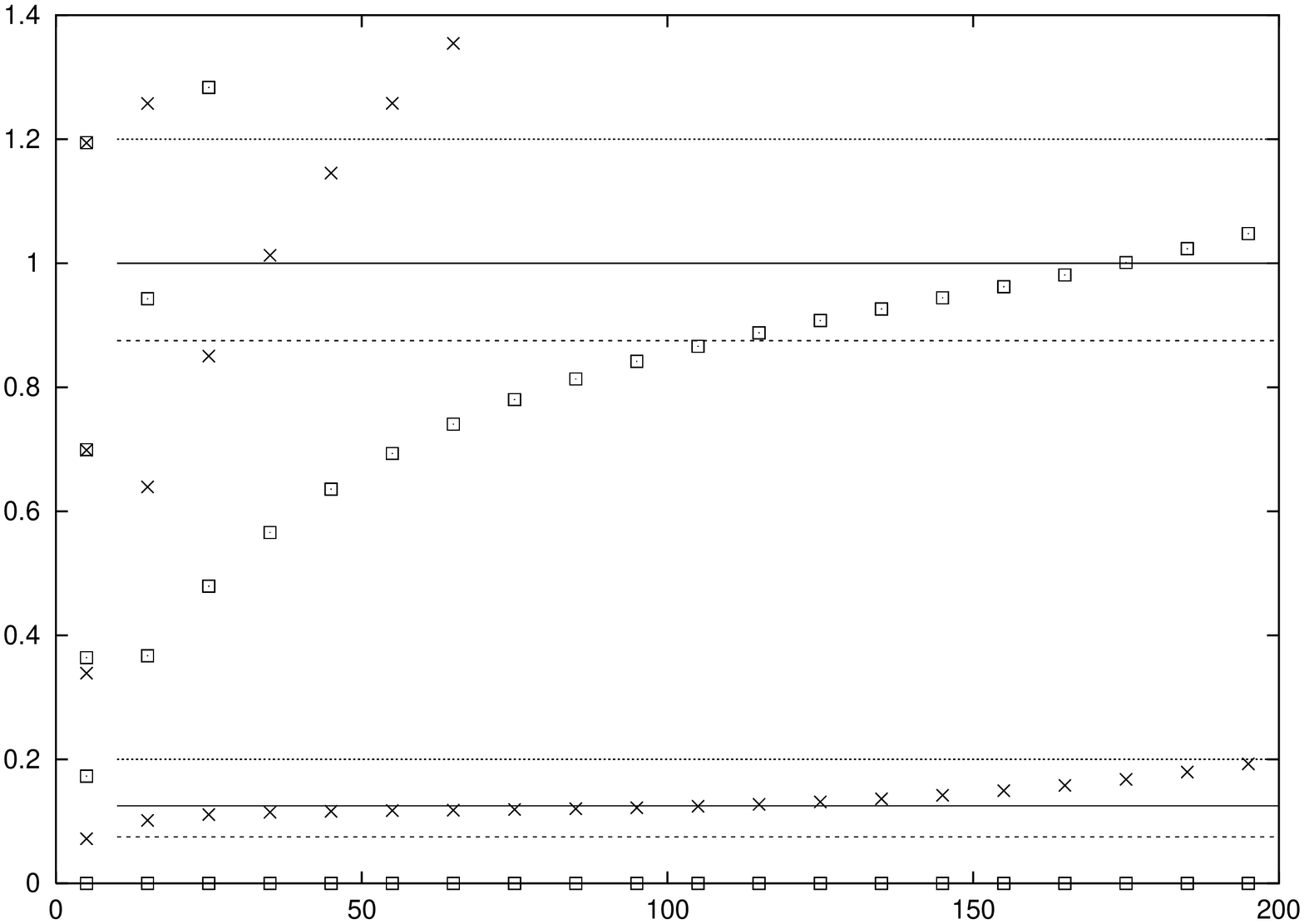}\includegraphics[%
  width=0.17\paperwidth,
  height=0.40\paperwidth,
  angle=270]{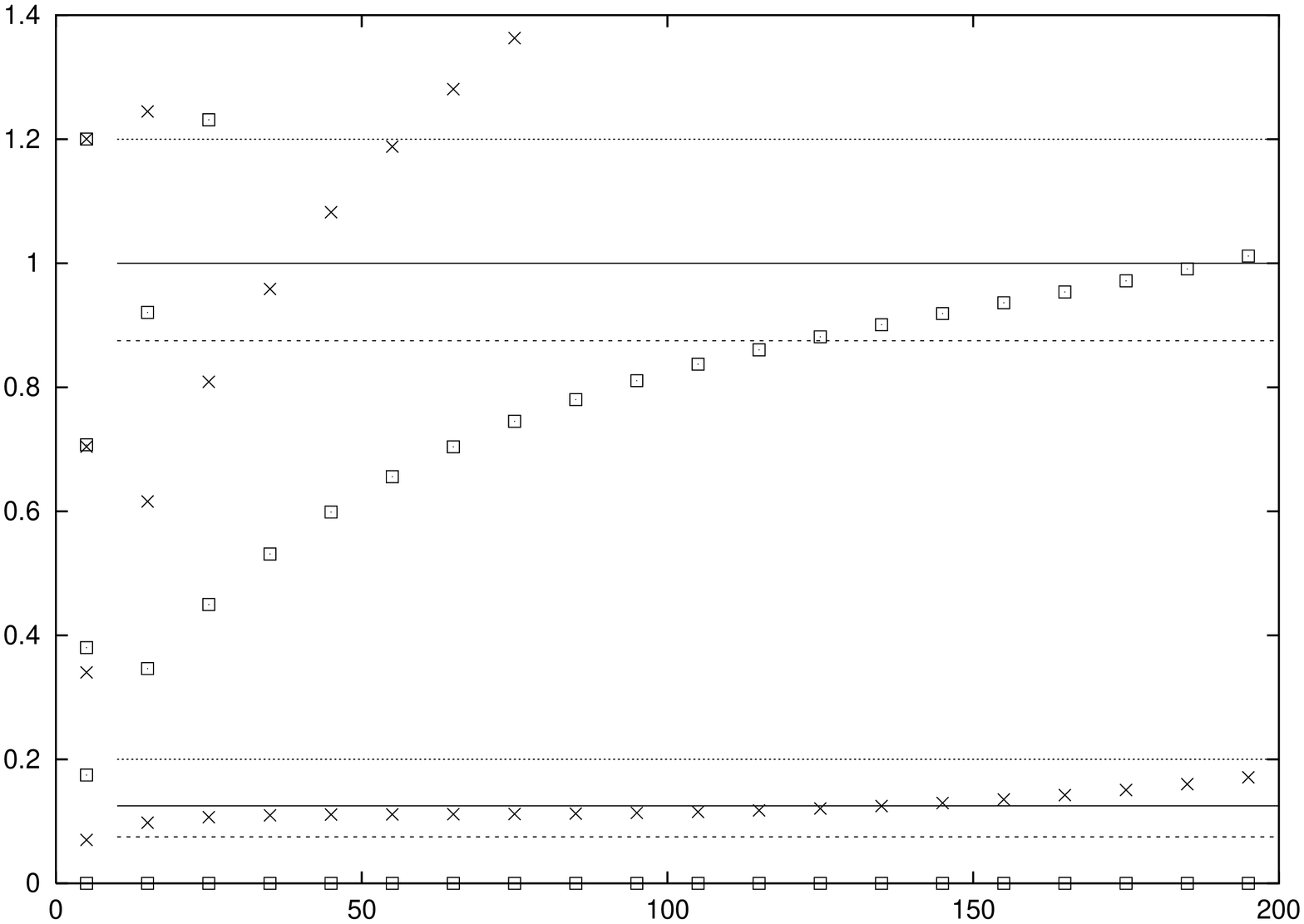}\end{center}

\vspace{-0.4cm} \hspace{3.5cm}a: $\eta_{1}=0.01$\hspace{6cm}b:
$\eta_{1}=0.05$ \vspace{-1cm}

\begin{center}\includegraphics[%
  width=0.17\paperwidth,
  height=0.40\paperwidth,
  angle=270]{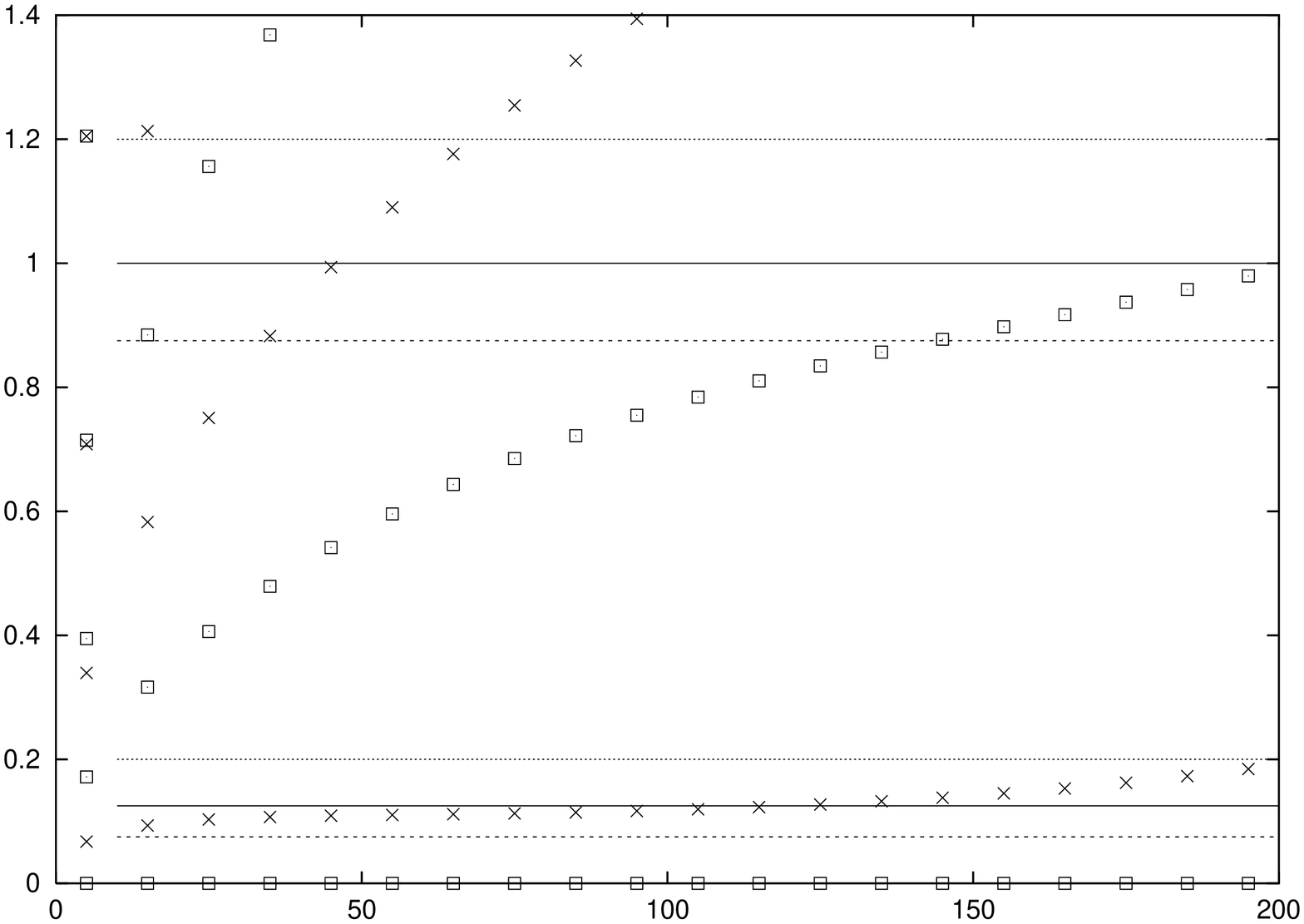}\includegraphics[%
  width=0.17\paperwidth,
  height=0.40\paperwidth,
  angle=270]{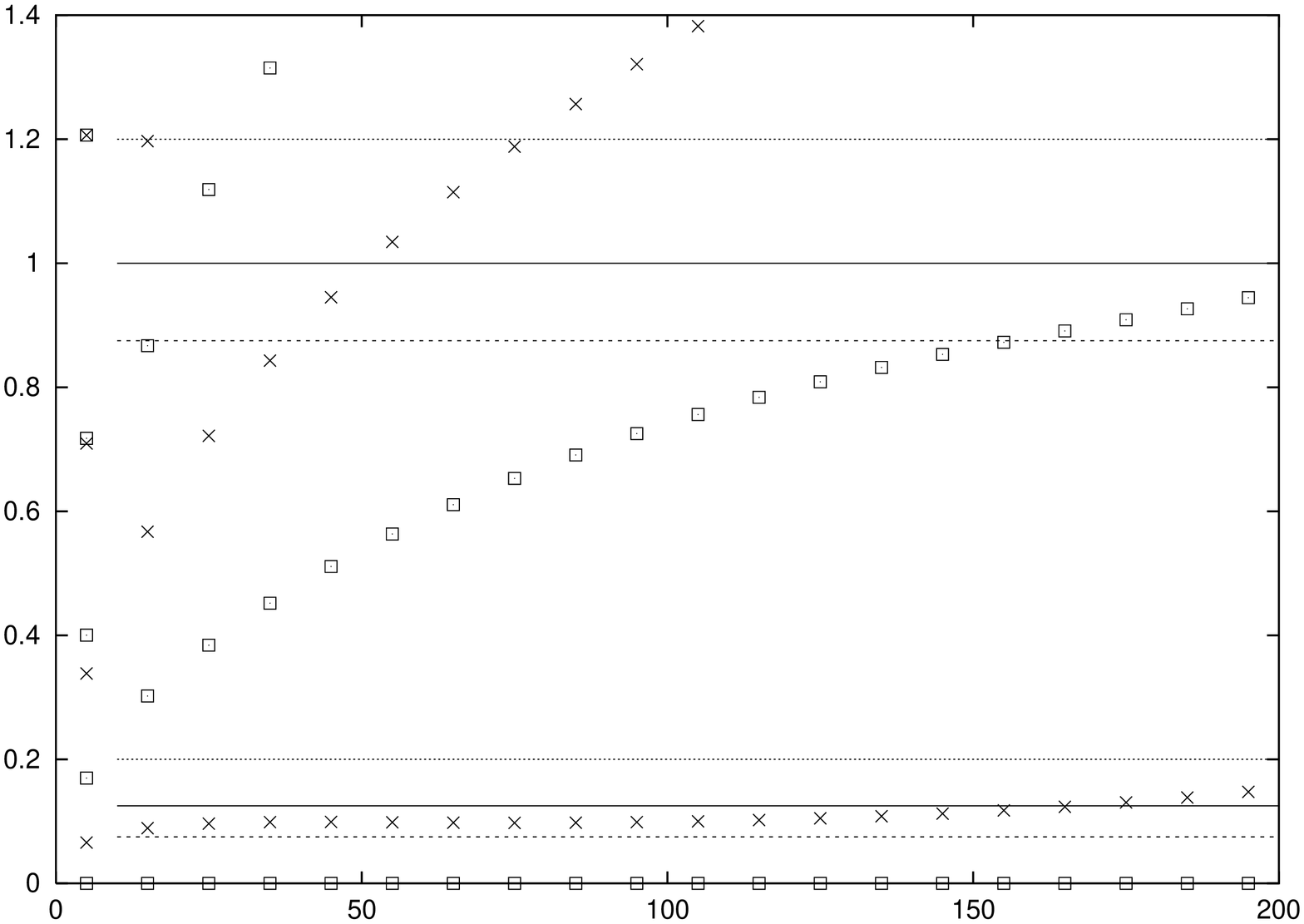}\end{center}

\vspace{-0.4cm} \hspace{3.5cm}c: $\eta_{1}=0.08$\hspace{6cm}d:
$\eta_{1}=0.09$ \vspace{-1cm}

\begin{center}\includegraphics[%
  width=0.17\paperwidth,
  height=0.40\paperwidth,
  angle=270]{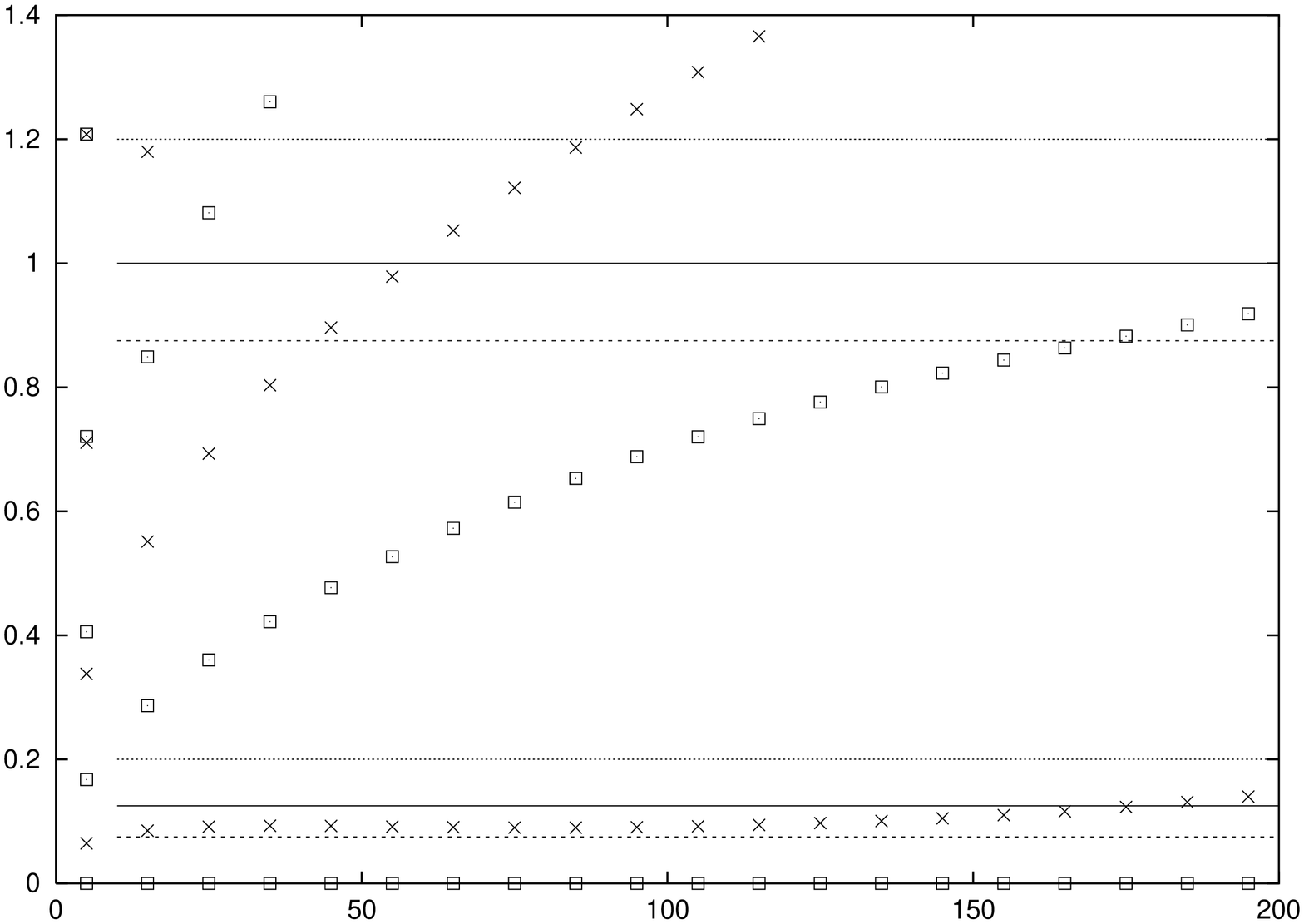}\includegraphics[%
  width=0.17\paperwidth,
  height=0.40\paperwidth,
  angle=270]{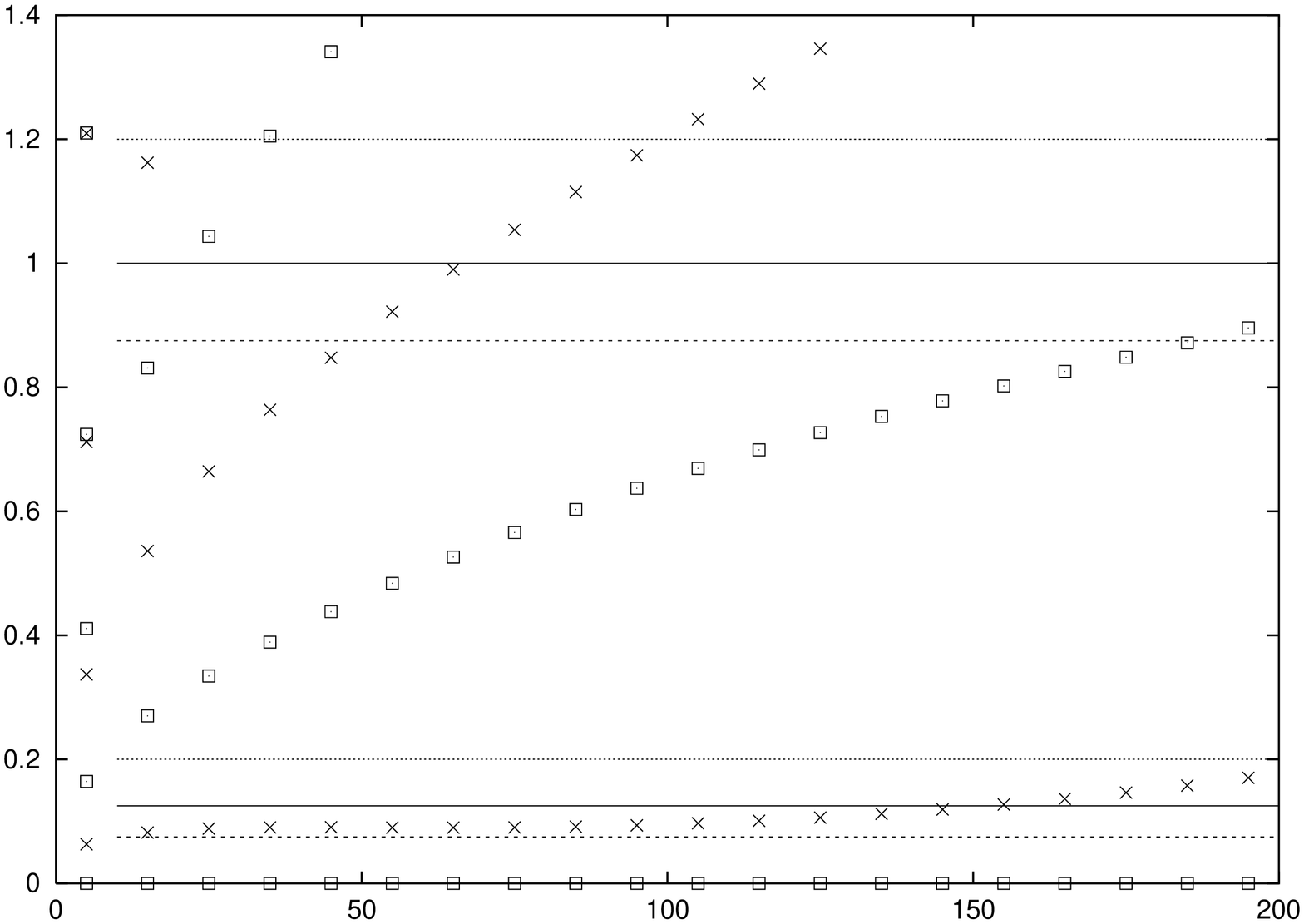}\end{center}

\vspace{-0.4cm} \hspace{3.5cm}e: $\eta_{1}=0.10$\hspace{6cm}f:
$\eta_{1}=0.11$ \vspace{-1cm}

\begin{center}\includegraphics[%
  width=0.17\paperwidth,
  height=0.40\paperwidth,
  angle=270]{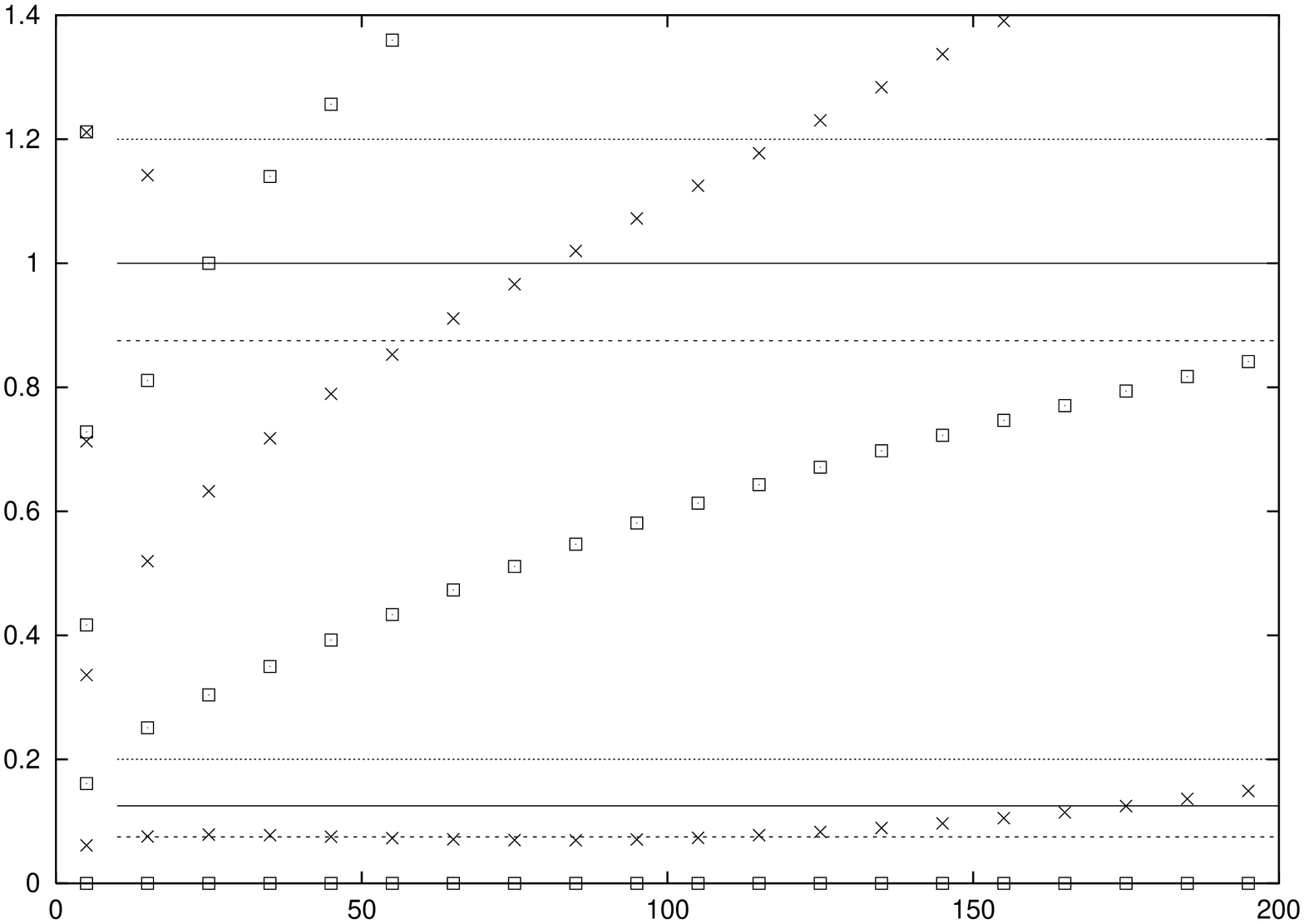}\includegraphics[%
  width=0.17\paperwidth,
  height=0.40\paperwidth,
  angle=270]{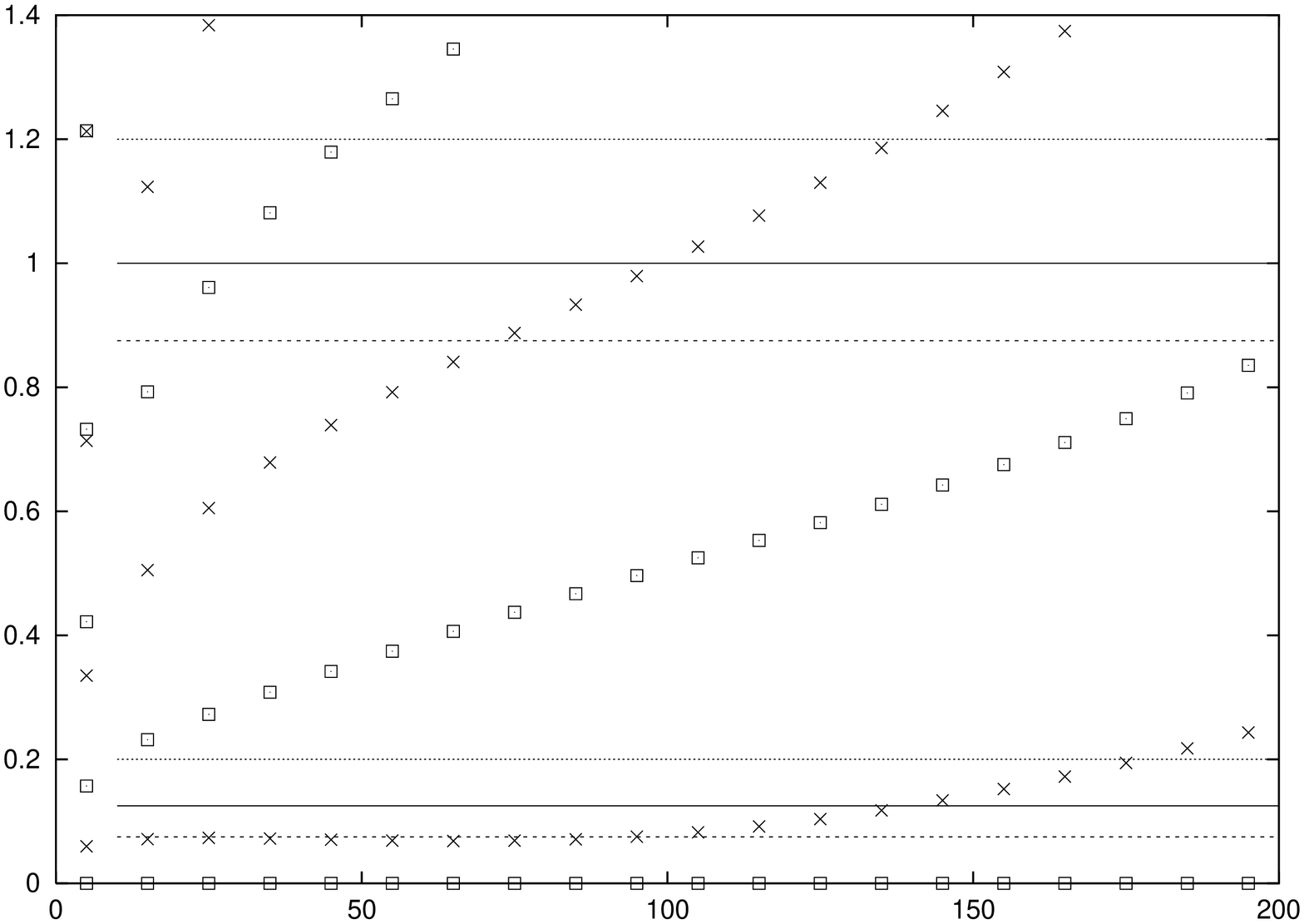}\end{center}

\vspace{-0.4cm} \hspace{3.5cm}g: $\eta_{1}=0.12$\hspace{6cm}h:
$\eta_{1}=0.13$
\end{figure}
\noindent Moreover, the value of $l$ where the truncation errors become large
gets smaller and smaller as the tricritical point is approached, which
corresponds to the fact that a tricritical point as renormalization
group fixed point is more repelling than an Ising type one. Increasing
$\eta_{1}$ further the behaviour corresponding to the tricritical
point would rapidly disappear from the finite volume spectrum.

\subsection{The line of first order transition}

\begin{figure}[H]
\begin{center}\includegraphics[%
  width=0.17\paperwidth,
  height=0.40\paperwidth,
  angle=270]{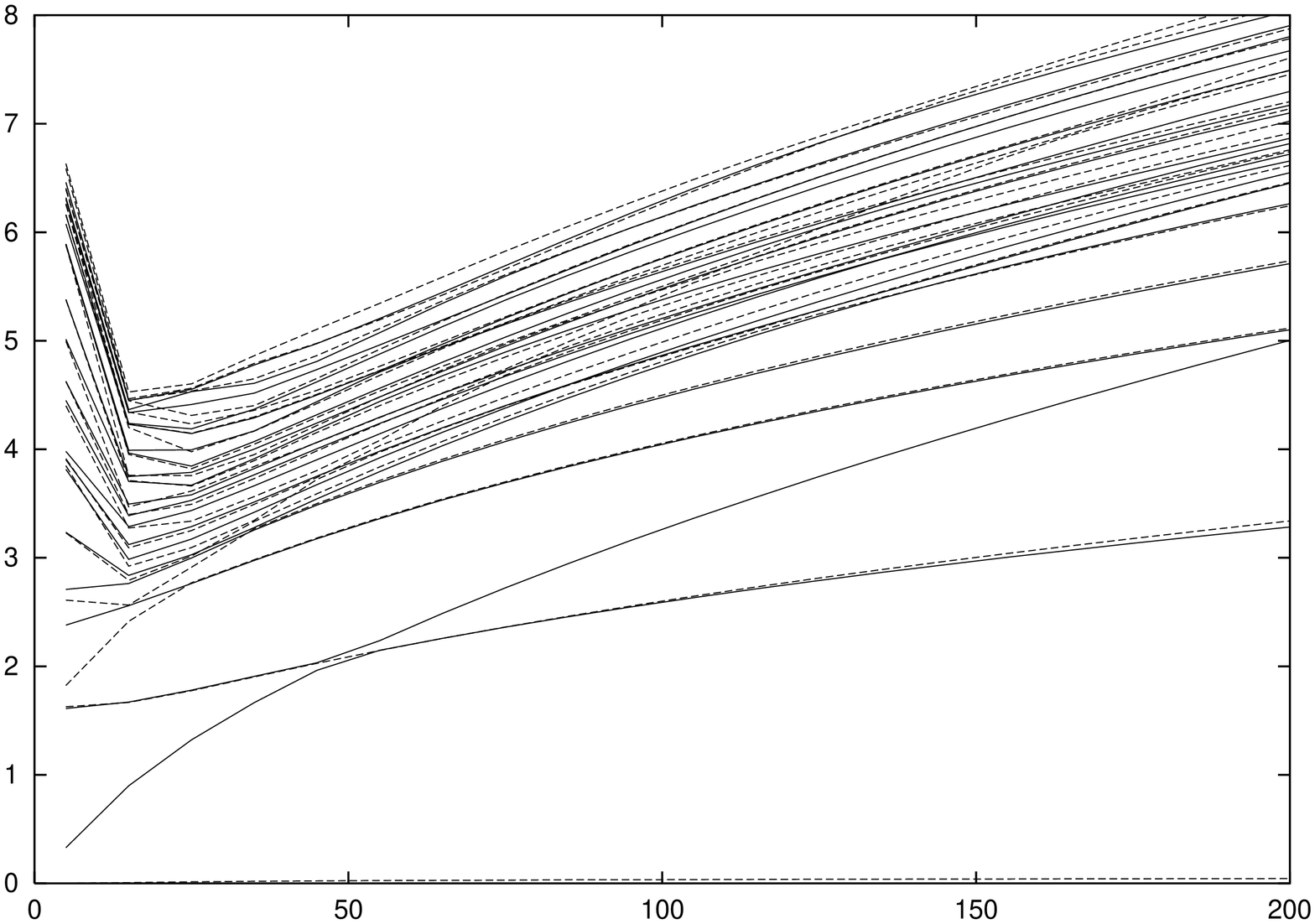}\includegraphics[%
  width=0.17\paperwidth,
  height=0.40\paperwidth,
  angle=270]{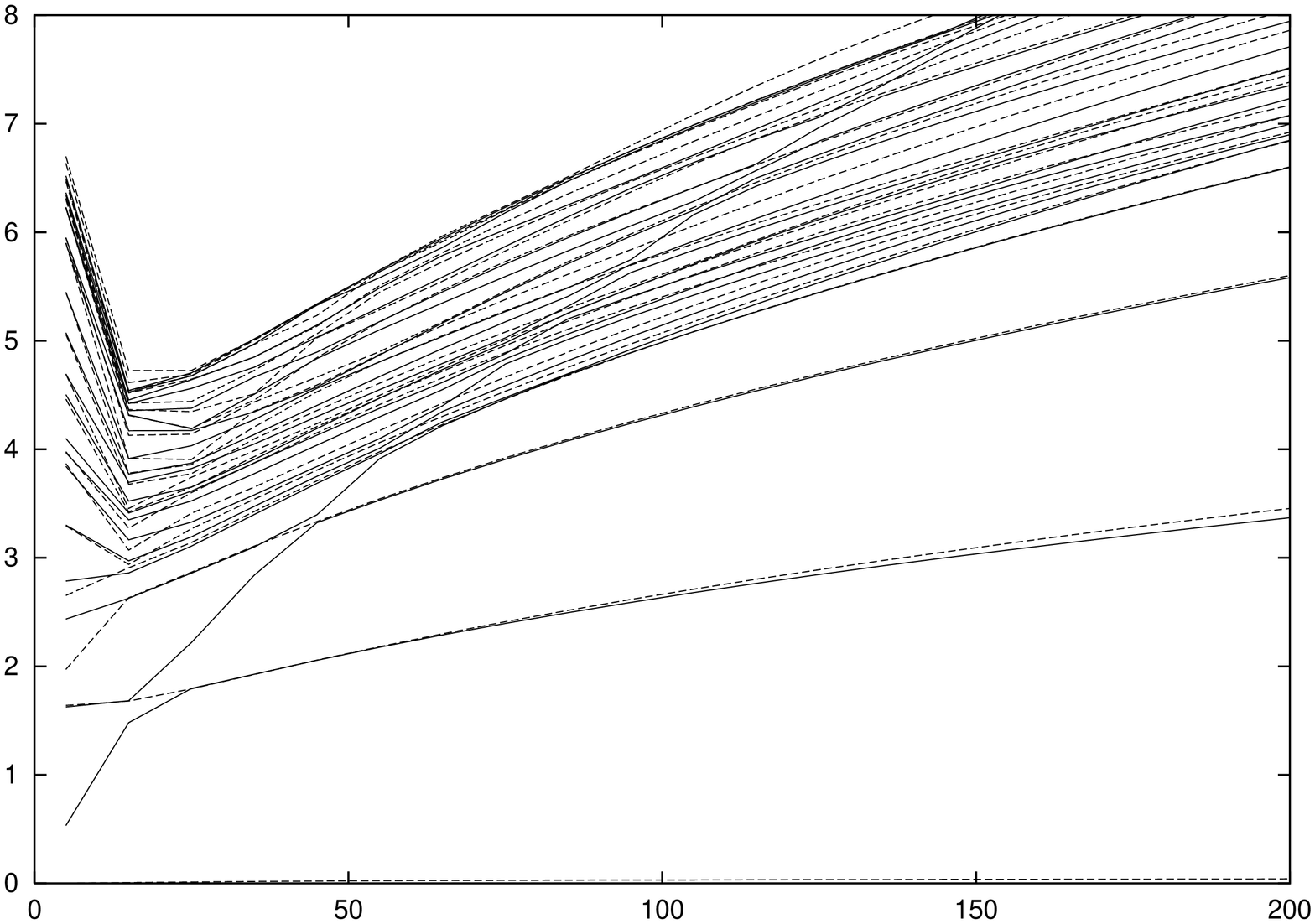}\end{center}

\vspace{-0.4cm} \hspace{3.5cm}i: $\eta_{2}=0.25$\hspace{6cm}j:
$\eta_{2}=0.30$

\caption{\label{nemkrit atmenet}$[e_{i}(l)-e_{0}(l)]$, $i=0..19$ as functions
of $l$ obtained by TCSA at $\beta=8\sqrt{\pi}/5$, $\eta_{1}=0.6$
and at various values of $\eta_{2}$ }
\end{figure}

Figures \ref{nemkrit atmenet}.a-j show TCSA spectra obtained at $\eta_{1}=0.6$
and at various values of $\eta_{2}$ between $0.17$ and $0.30$.
The energy levels are shown as compared to the lowest level. The dimension
of the truncated Hilbert space was $6597$ in these calculations,
which corresponded to $e_{cut}=16$. The value of $\beta$ was $8\sqrt{\pi}/5$.
Dashed lines are used for odd parity levels and continuous lines for
even parity levels.

At $\eta_{2}=0.17$ (and also for $\eta_{2}<0.17$) the ground state
is unique for all values of $l$, whereas doubly degenerate runaway
levels can also be seen. At $\eta_{2}=0.19$, however, the ground
level is doubly degenerate for small values of the volume but becomes
nondegenerate in large volume, %
\begin{figure}[H]
\begin{center}\includegraphics[%
  width=0.17\paperwidth,
  height=0.40\paperwidth,
  angle=270]{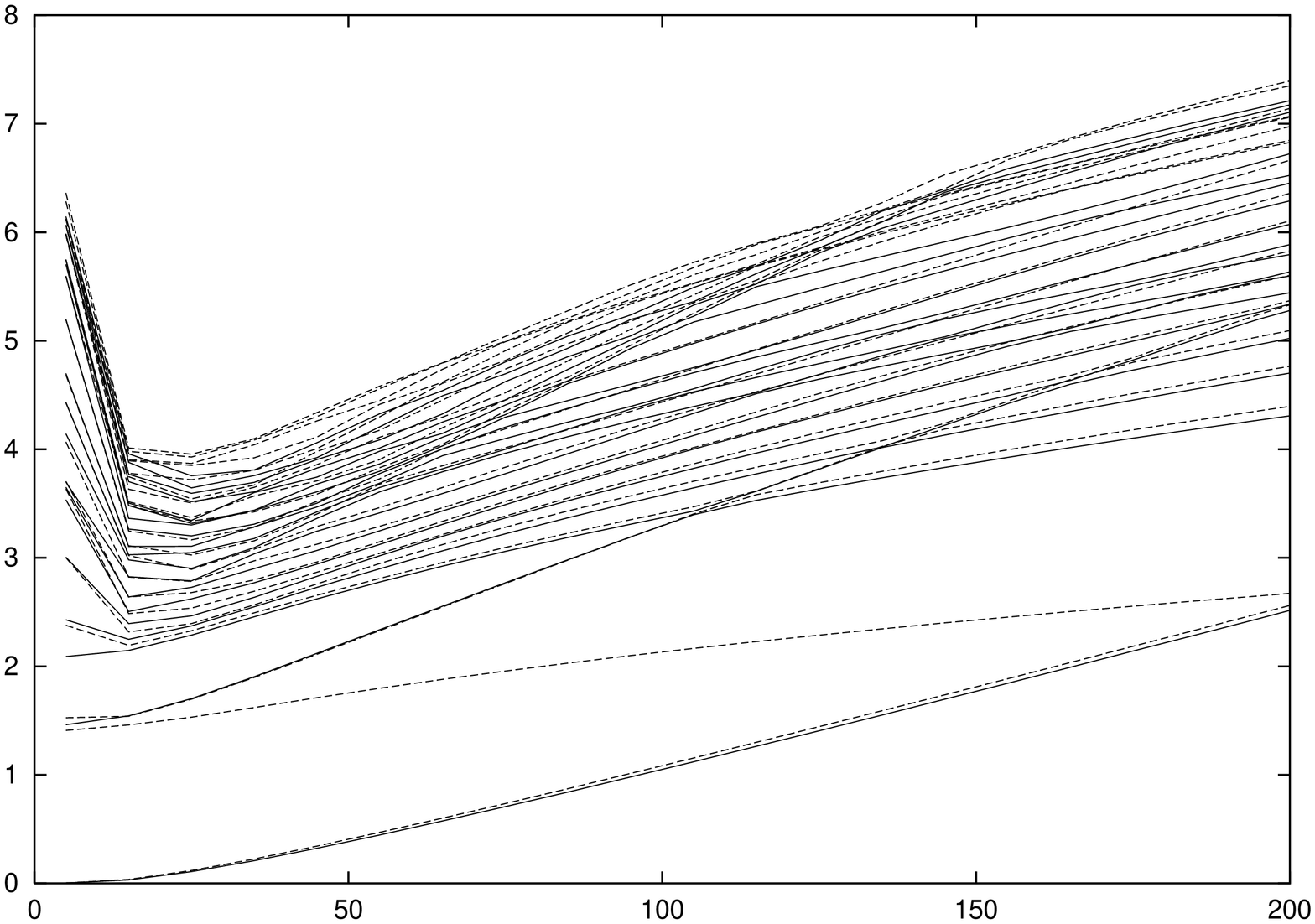}\includegraphics[%
  width=0.17\paperwidth,
  height=0.40\paperwidth,
  angle=270]{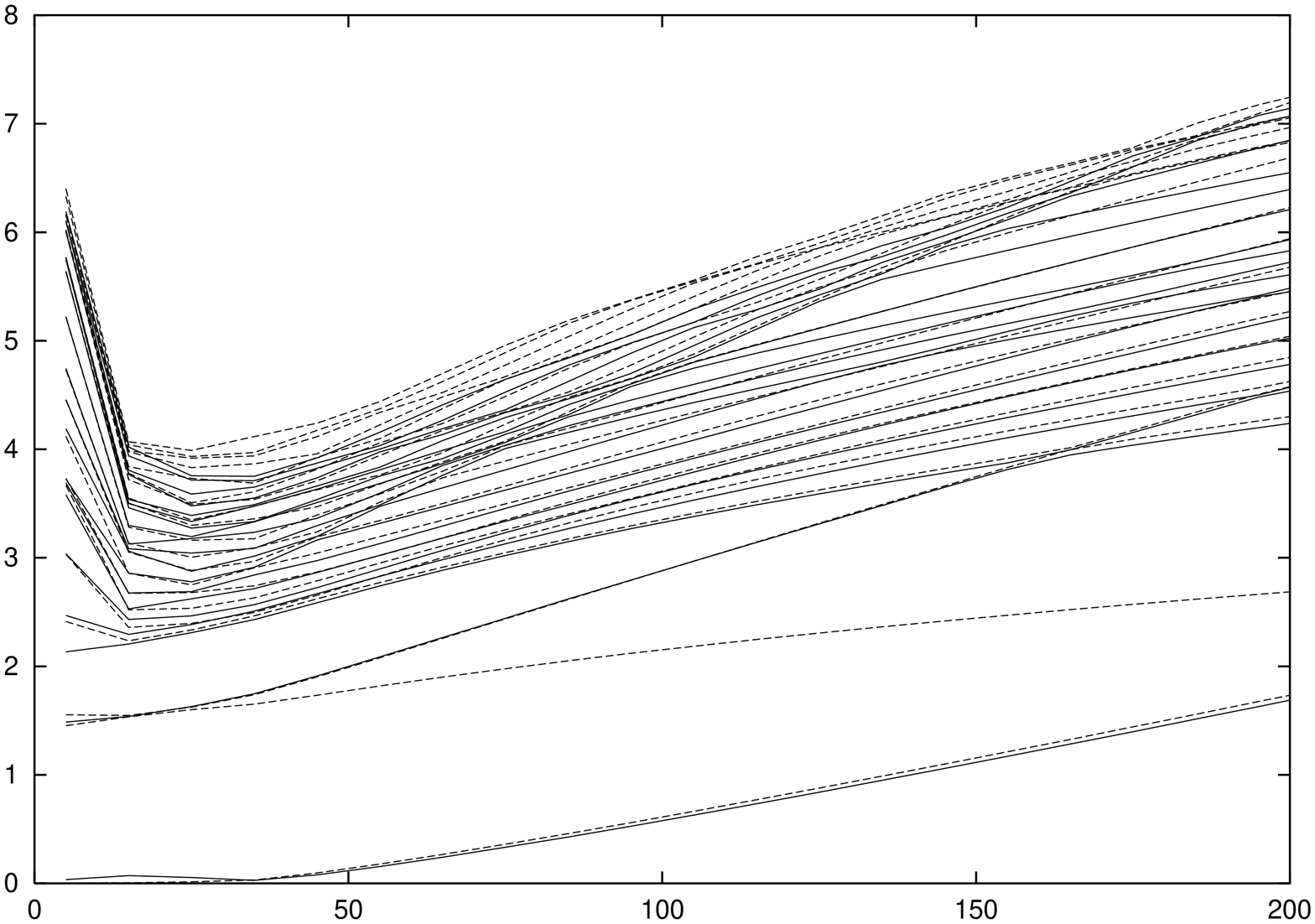}\end{center}

\vspace{-0.4cm} \hspace{3.5cm}a: $\eta_{2}=0.17$\hspace{6cm}b:
$\eta_{2}=0.18$ \vspace{-1cm}

\begin{center}\includegraphics[%
  width=0.17\paperwidth,
  height=0.40\paperwidth,
  angle=270]{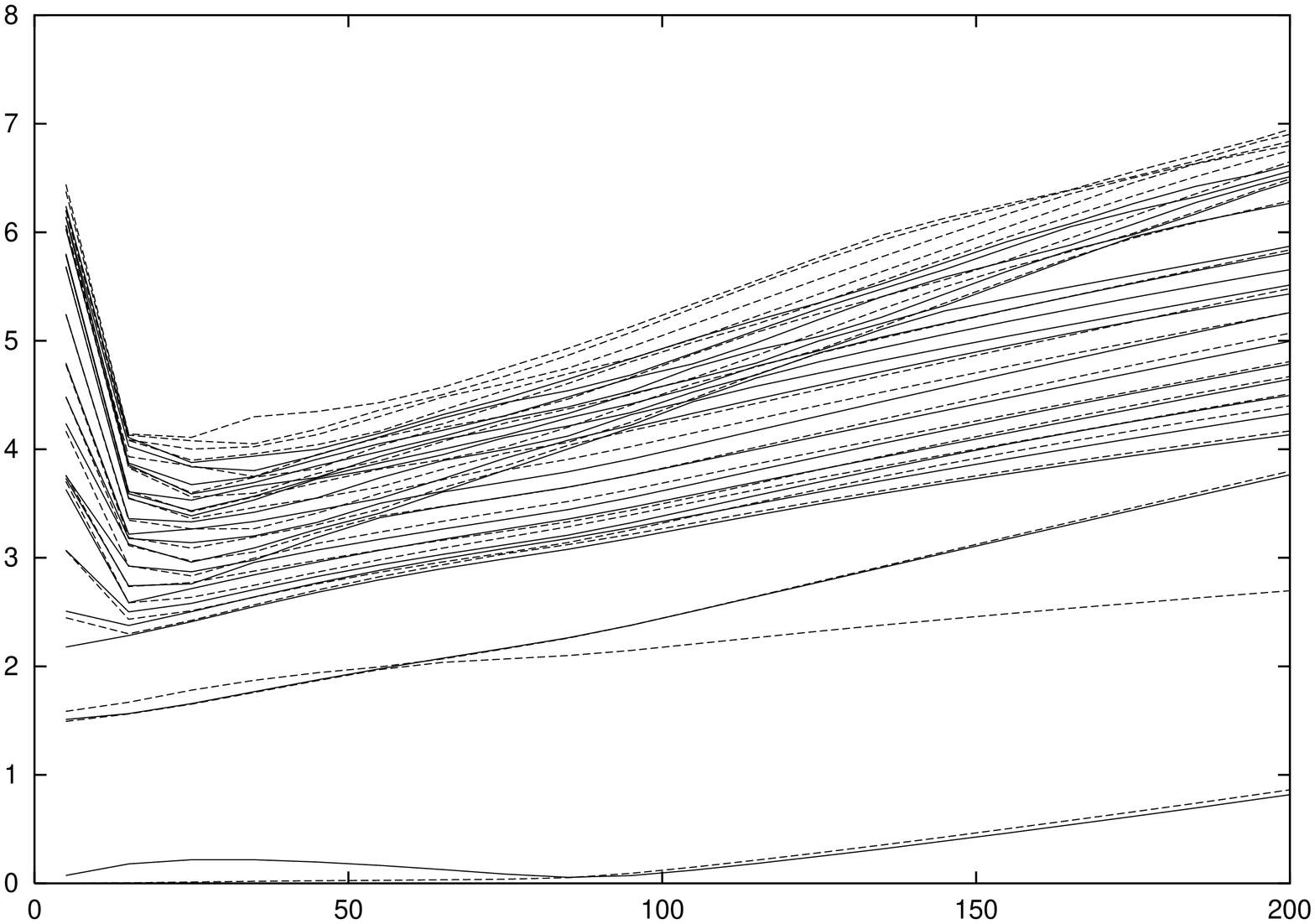}\includegraphics[%
  width=0.17\paperwidth,
  height=0.40\paperwidth,
  angle=270]{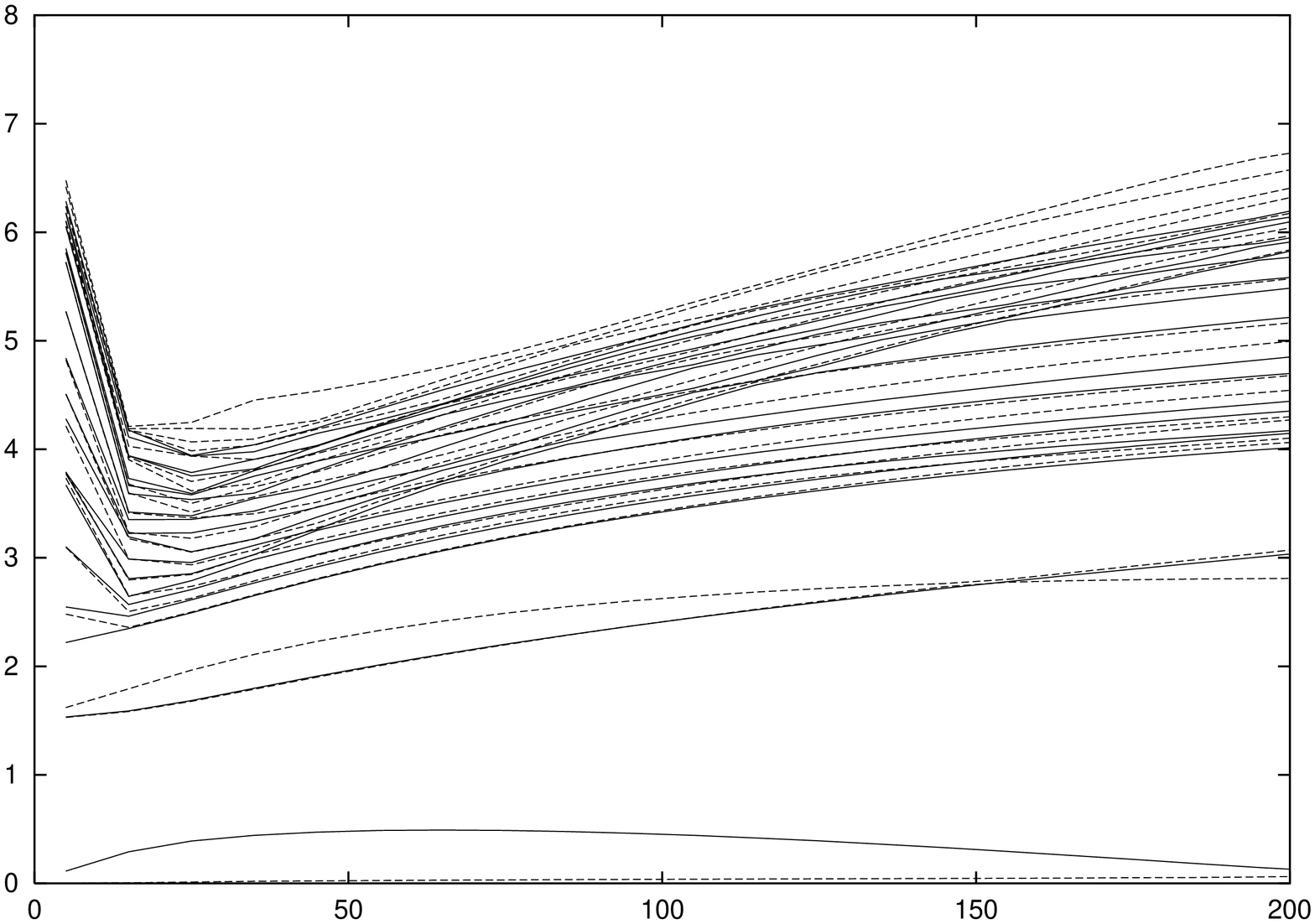}\end{center}

\vspace{-0.4cm} \hspace{3.5cm}c: $\eta_{2}=0.19$\hspace{6cm}d:
$\eta_{2}=0.20$ \vspace{-1cm}

\begin{center}\includegraphics[%
  width=0.17\paperwidth,
  height=0.40\paperwidth,
  angle=270]{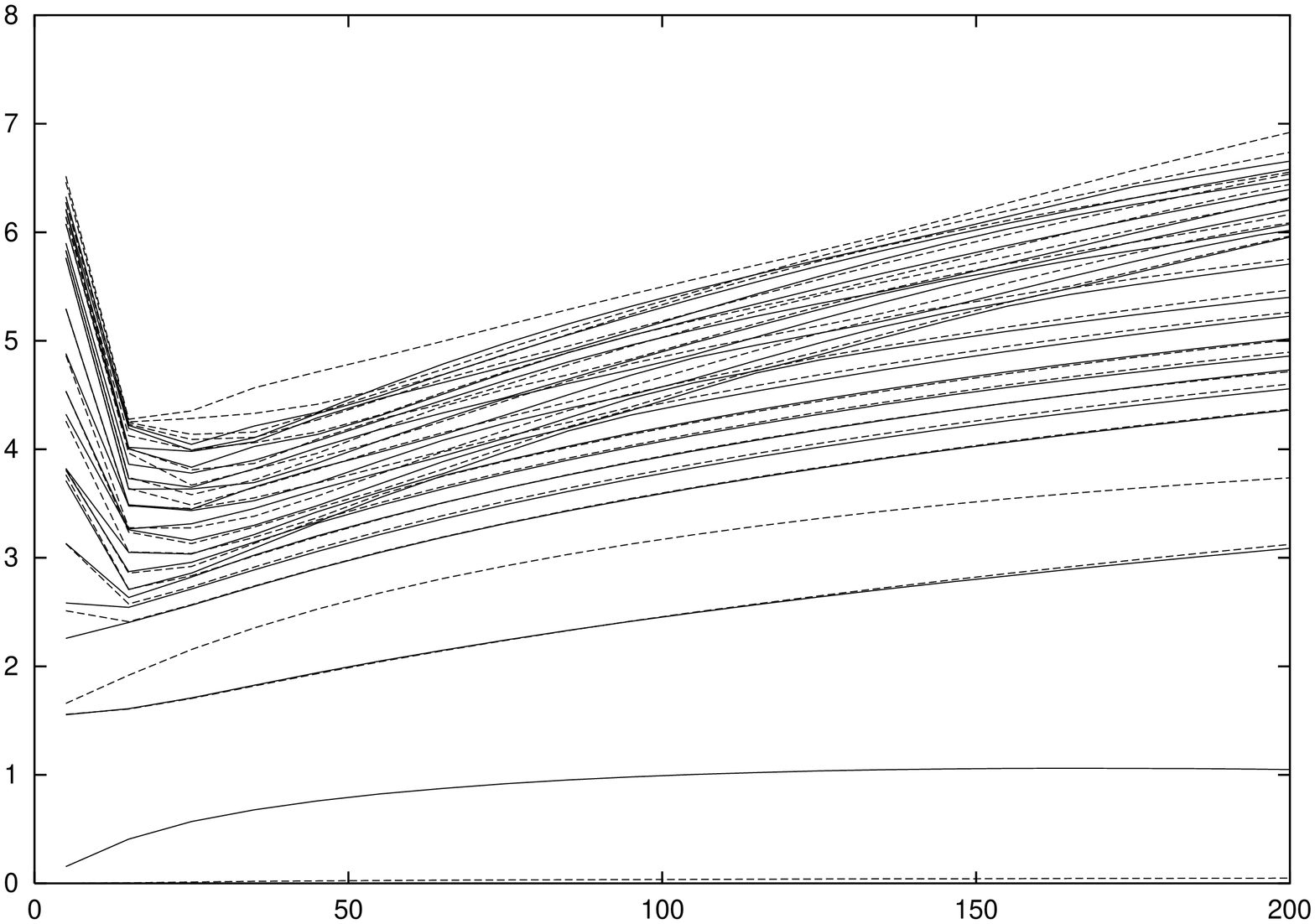}\includegraphics[%
  width=0.17\paperwidth,
  height=0.40\paperwidth,
  angle=270]{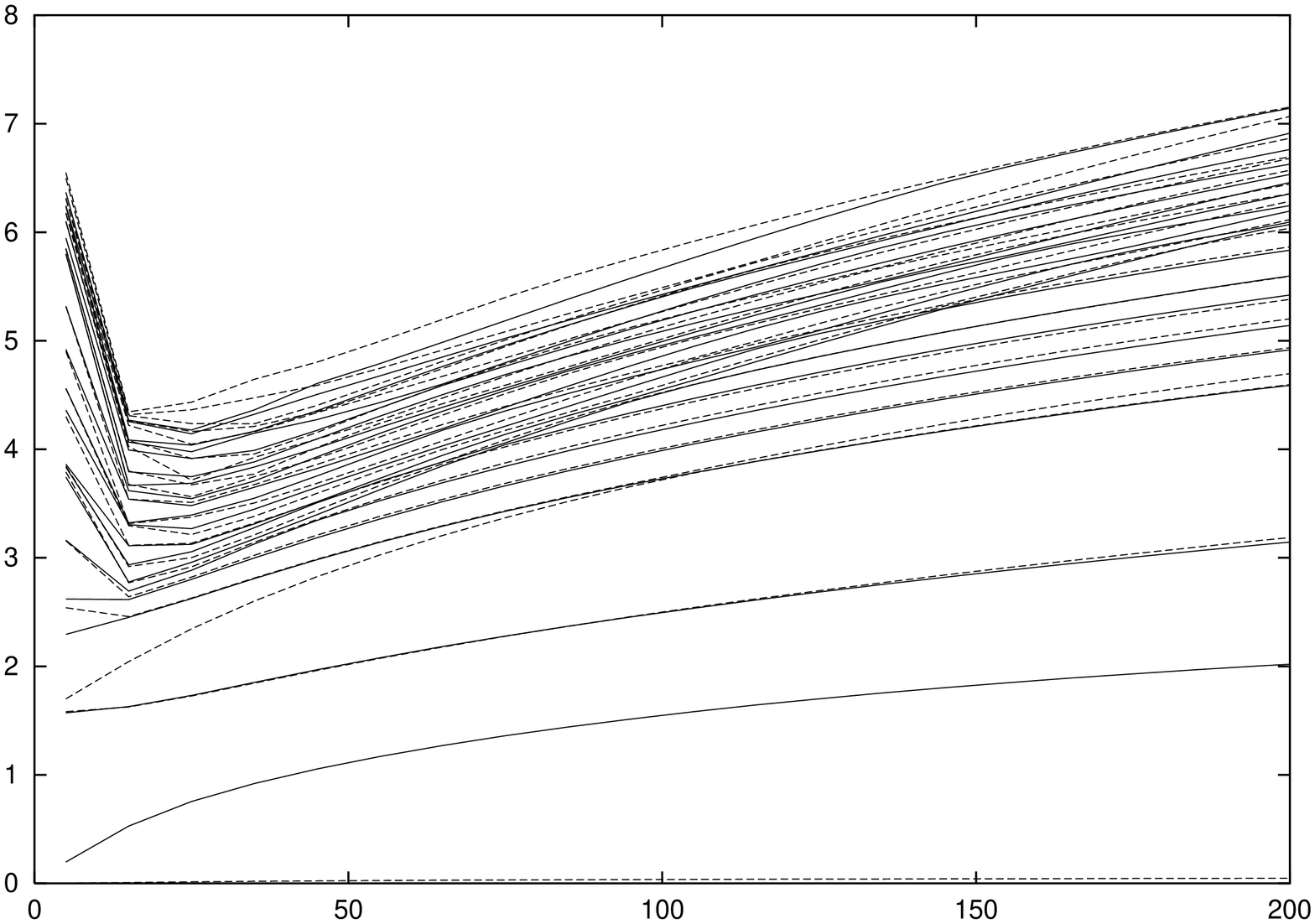}\end{center}

\vspace{-0.4cm} \hspace{3.5cm}e: $\eta_{2}=0.21$\hspace{6cm}f:
$\eta_{2}=0.22$ \vspace{-1cm}

\begin{center}\includegraphics[%
  width=0.17\paperwidth,
  height=0.40\paperwidth,
  angle=270]{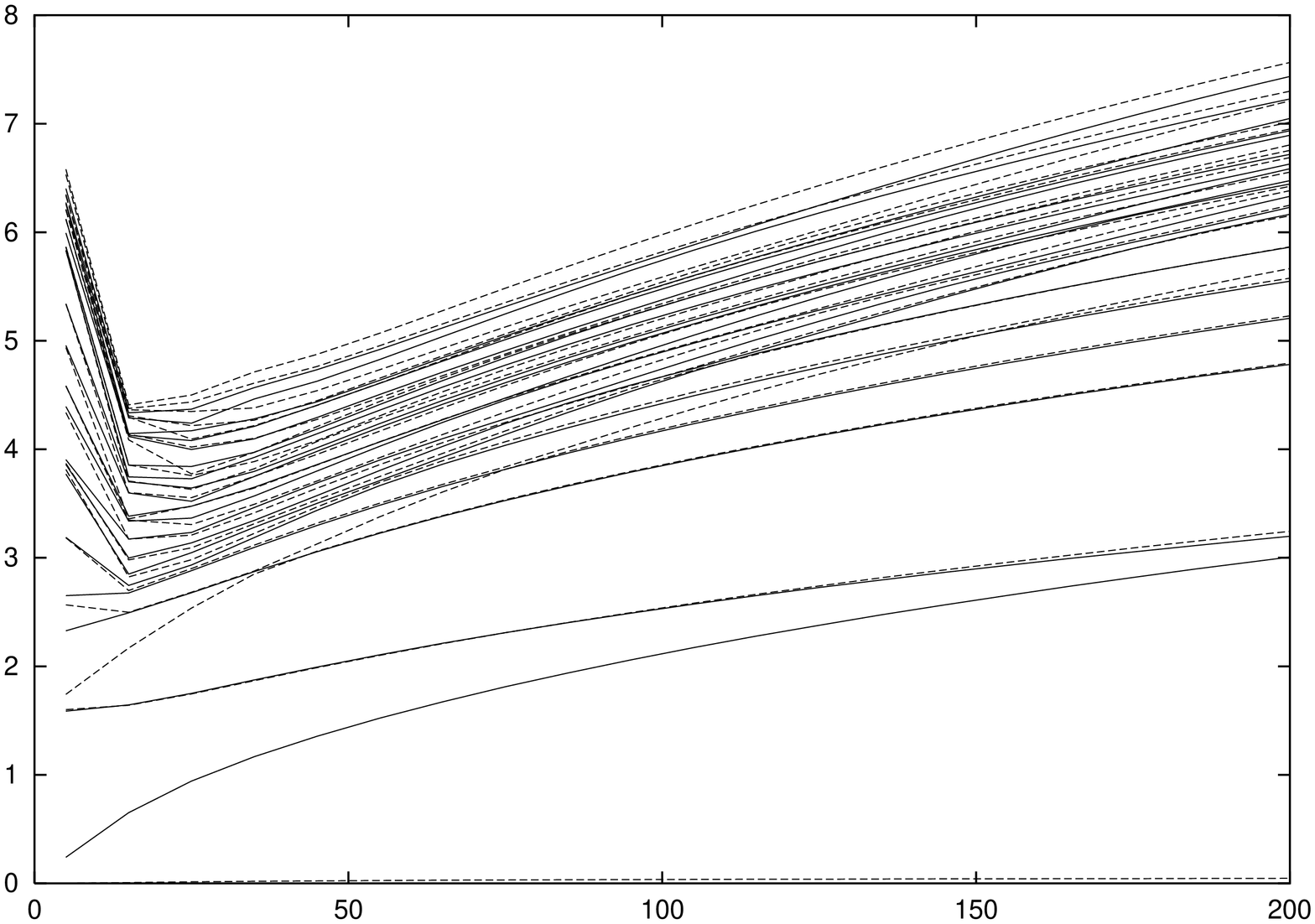}\includegraphics[%
  width=0.17\paperwidth,
  height=0.40\paperwidth,
  angle=270]{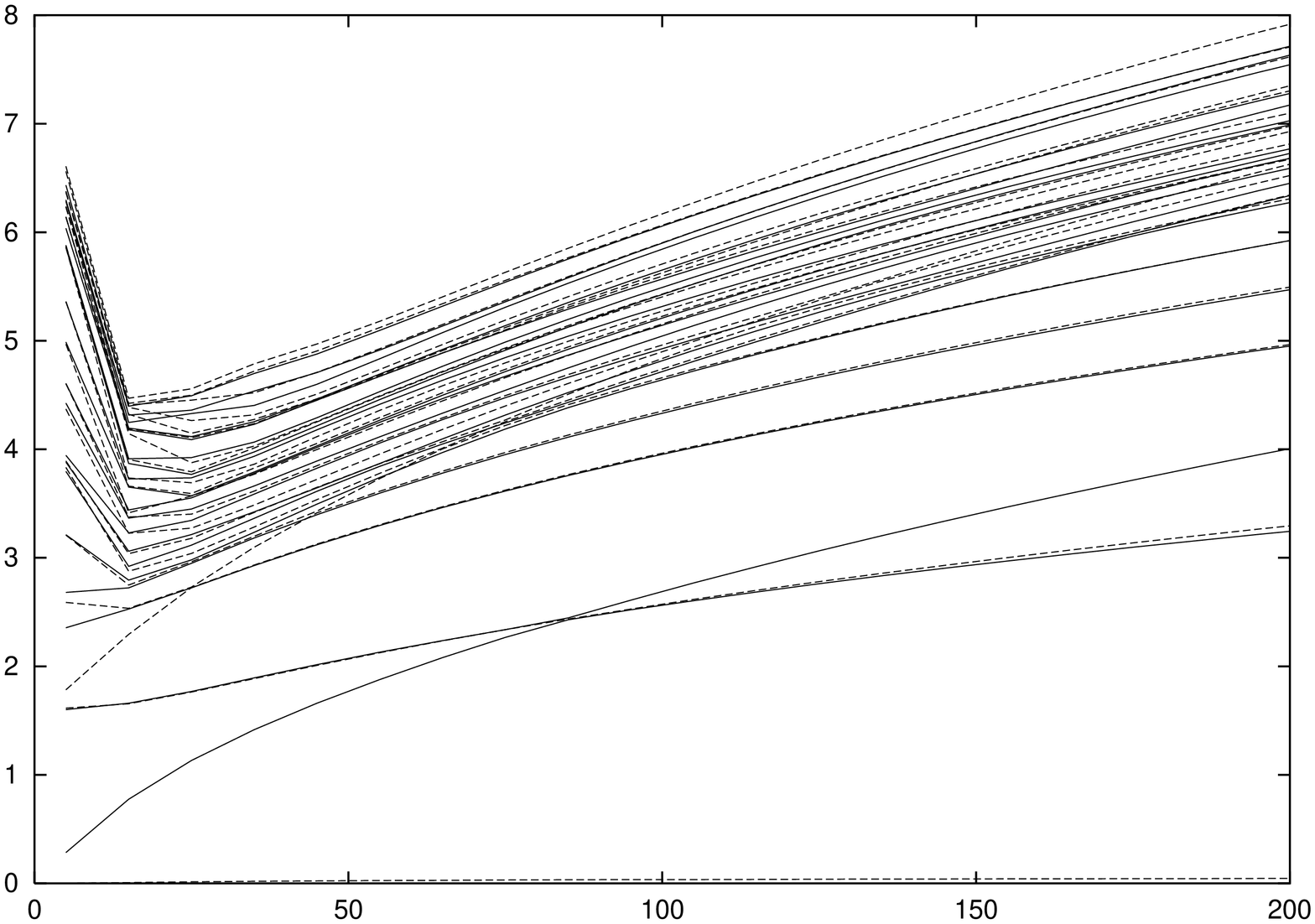}\end{center}

\vspace{-0.4cm} \hspace{3.5cm}g: $\eta_{2}=0.23$\hspace{6cm}h:
$\eta_{2}=0.24$
\end{figure}
\noindent i.e. there is a value of $l$ where the doubly degenerate runaway
level and the single level cross each other. The slope of the runaway
levels become smaller and smaller as $\eta_{2}$ is increased, and
the crossing point also moves towards larger and larger values. At
$\eta_{2}=0.3$ the ground state is already doubly degenerate for
all values of $l$ and nondegenerate runaway levels are present. These
features indicate that a first order phase transition occurs at an
intermediate value of $\eta_{2}$. The behaviour of the finite volume
spectra around the transition point seen in the figures implies that
a precise determination of the transition point from the TCSA data
in a direct way would require precise data for large values of $l$.
For this reason we did not aspire to find many points of the line
of first order phase transition, we did TCSA calculations only at
$\eta_{1}=0.4$ and $\eta_{1}=0.6$, and we roughly estimated the
location of the first order phase transition at these values. Our
estimation based on the TCSA data are $\eta_{2}=0.30$ and $\eta_{2}=0.21$.
These points are also marked in Figure \ref{threefreq phasediag}
by crosses. The Figures \ref{nemkrit atmenet}.a-j show the first
$20$ levels in the domain $l=0..200$. Unfortunately the truncation
effect is large in most of this domain of $l$, however we expect
that those qualitative features of the spectra which we allude to
are correct.

\section{\label{sec:Conclusions}Conclusions}

In this work we investigated selected special cases of the multi-frequency
sine-Gordon theory, especially the structure of their phase space, continuing
the work begun in \cite{BPTW}. Concerning the classical limit we
found that the only possible phase transition in the (rational) two-frequency
case is a second order Ising-type transition. The three-frequency
model has some new qualitative features compared to the two-frequency
model: a tricritical point which is at the end of a critical line
can also be found, and first order transition is possible as well.
The phase space of the $n$-frequency model contains $n$-fold critical
points, otherwise we do not expect new qualitative features compared
to the three-frequency model.

The numerical (TCSA) calculations in the quantum case yielded the
following results: we found the same type of phase transitions as
in the classical case, in particular we were able to determine the
second order Ising nature and the location of the phase transition
in the two-frequency model with good precision. Quantum corrections
did not alter the nature of the phase transition. The accuracy of
the TCSA also allowed us to find the tricritical point in the three-frequency
case, which we regard the main result of the paper. This demanded
much more numerical work than the two-frequency model for the following
reasons: the two-frequency model has only one-dimensional phase space,
whereas the phase space is two-dimensional in the three-frequency
model; and the tricritical point as a renormalization group fixed
point is more repelling than the Ising-type critical point, so we
had to take larger truncated space. Furthermore, we found several
points of the critical line and observed how the TCSA spectrum changes
as the tricritical endpoint is approached. It would be interesting
to investigate this with considerably better precision. Although the
focus was mainly on the tricritical point, we also investigated the
first order phase transition in the three-frequency model and we found
that the first order nature can be established by TCSA, but the location
of the transition can be determined much less accurately than that
of the second-order transition. We expect that multicritical points
in the universality classes of further elements of the discrete unitary
series could be found in the higher-frequency models, but increasing
accuracy is needed, because these multicritical points are more and
more repelling. Finally, we remark that the quantum corrections did
not alter the nature of the phase transitions in any of the cases
we investigated, and in the classical and quantum theory the location
of transition points is almost the same. The investigation of the
multi-frequency model with irrational frequency ratios is still an
open problem. We also remark that the particle content of the multi-frequency
sine-Gordon model could also be investigated by the method of semiclassical
quantization \cite{key-23}.

\subsection*{Acknowledgments}

I would like to thank L. Palla, Z. Bajnok, G. Tak\'{a}cs
and F. W\'{a}gner for useful discussions and for their TCSA program.
I would also like to thank the Mathematics 
Department, King's College London, and especially  Gerard Watts and Andreas Recknagel 
for their kind hospitality.

This work was partially supported by the Hungarian fund OTKA T037674. 
I also thank  
the EU Research Training Network ``EUCLID'', contract HPRN-CT-2002-00325, 
for support at 
various times.
I also acknowledge the support by the EU 
Marie Curie Training Site MCFH-2001-00296 ``Strings, Branes and Boundary 
Conformal Field 
Theory'' at King's College London.

\end{document}